\documentclass[journal]{IEEEtran}
\usepackage{amsmath}
\usepackage{graphicx}
\usepackage{amssymb,amsmath}
\usepackage{mathtools}
\usepackage{autonum}
\usepackage[usestackEOL]{stackengine}
\usepackage[normalem]{ulem}
\usepackage{balance}

\usepackage{subfigure}
\usepackage{graphicx,epsfig}
\usepackage[linesnumbered,ruled,vlined]{algorithm2e} 
\usepackage{empheq}
\usepackage{cite}
\usepackage{color}
\usepackage[normalem]{ulem}

\newtheorem{theorem}{Theorem}

\newcommand{\mt}[1]{}
\let \mt=\mathrm

\newcounter{MYtempeqncnt}

\makeatletter

\newcommand{\Rmnum}[1]{\expandafter\@slowromancap\romannumeral #1@}

\makeatletter

\newcommand{\comm}[1]{}

\let \bd = \textbf
\let \bs = \boldsymbol

\let \it = \textit
\let \t  = \text

\def \p1#1{#1^{-1}}

\def \~#1{\tilde{#1}}

\ifCLASSINFOpdf
\else
\fi
\begin{document}
%
\title{Multi-stage Antenna Selection for Adaptive Beamforming in MIMO Arrays}

\author{\IEEEauthorblockN{Hamed Nosrati, \IEEEmembership{Studnet Member, IEEE}, Elias Aboutanios, \IEEEmembership{Senior Member, IEEE}, and
David Smith, \IEEEmembership{Member, IEEE}}\\


\thanks{H. Nosrati is with the School of Electrical Engineering and
Telecommunication, University of New South Wales, NSW 2052, Australia, and Data61, CSIRO (Commonwealth Scientific and Industrial Research Organisation), NSW 2015, Australia    (e-mail:hamed.nosrati@unsw.edu.au).

E. Aboutanios is with the School of Electrical Engineering and
Telecommunication, University of New South Wales, NSW 2052, Australia
(e-mail:elias@unsw.edu.au)

D. Smith, is with Data61, CSIRO (Commonwealth Scientific and Industrial Research Organisation), NSW 2015, Australia 
(e-mail:david.smith@data61.csiro.au)}

}


%


\maketitle
\begin{abstract}
Increasing the number of transmit and receive elements in multiple-input-multiple-output (MIMO) antenna arrays imposes a substantial increase in hardware and computational costs. We mitigate this problem by employing a reconfigurable MIMO array where large transmit and receive arrays are multiplexed in a smaller set of $k$ baseband signals. We consider four stages for the MIMO array configuration and propose four different selection strategies to offer dimensionality reduction in post-processing and achieve hardware cost reduction in digital signal processing (DSP) and radio-frequency (RF) stages. 
We define the problem as a determinant maximization and develop a unified formulation to decouple the joint problem and select antennas/elements in various stages in one integrated problem.
We then analyze the performance of the proposed selection approaches and prove that, in terms of the output SINR, a joint transmit-receive selection method performs best followed by matched-filter, hybrid and factored selection methods. The theoretical results are validated numerically, demonstrating that all methods allow an excellent trade-off between performance and cost. 
\end{abstract}
\begin{IEEEkeywords}
Antenna selection, MIMO radar, adaptive array beamforming, STAP, convex optimization.
\end{IEEEkeywords}



%
\IEEEpeerreviewmaketitle
\section{Introduction}\label{intro}
   
The spatial diversity and performance improvements offered by multiple-input multiple-output (MIMO) antenna systems have led to their widespread use in a variety of applications including wireless communications e.g. massive MIMO \cite{Larsson2014a,Mietzner2009}, radar and sonar \cite{Li,Hassanien2010}. In radar, MIMO arrays have proven effective at enhancing the radar's resolution as they offer increased number of degrees of freedom (DOFs) \cite{Bliss2003}. A MIMO phased array comprises an array of antennas, transmitting a set of noncoherent orthogonal waveforms that can be extracted at the receiver by a corresponding number of matched filters. Improved spatial diversity, parameter identifiability, and detection performance result from the added DoFs compared to single-input multiple-output (SIMO) configurations~\cite{Li}.

The advantages of the MIMO configuration are delivered at the expense of a significant increase in the problem dimensionality and hardware cost \cite{molisch2004mimo}. The system hardware include the antennas, baseband digital signal processing (DSP) and radio frequency (RF) front-ends comprising the low noise amplifiers (LNA), phase shifters, and frequency mixers. Among these, the baseband DSP and RF front-ends are a great deal more expensive than the antenna elements. One way to reduce the cost while maintaining the spectral diversity is to employ a large antenna array but select a subset of antennas to feed through the RF switching network \cite{Heath2001,Wang2014b,Wang2017}. In this work, we focus on antenna selection in the context of MIMO radar.

Over the last decade, antenna selection in MIMO arrays has commanded significant attention both in wireless communications and radar applications. In communications, antenna selection is employed to maximize the channel capacity. To this end, near-optimal strategies that assume perfect knowledge of the channel were proposed in \cite{Gharavi-Alkhansari2004} and \cite{Gorokhov2002}.  In \cite{Berenguer2005}, a fast adaptive antenna selection via discrete stochastic optimization in is proposed, where an aggressive stochastic approximation is employed to generate iteratively a sequence of estimates of the solution. More recently, antenna selection has been employed to reduce complexity and power consumption in mm-wave MIMO systems through compressed spatial sampling of the received signal \cite{Mendez-Rial2016}. In radar, both deterministic and optimization-based methods have been developed to select a subset of antennas and reconfigure the array architecture in order to maximize the output signal-to-interference and noise ratio (SINR) \cite{Liu2014,Keizer2008,Wang2014b,Wang2017} and enhance the direction of arrival (DoA) estimation \cite{Wang2015}. Antenna selection also plays an important role in aperture sharing in dual function radar communication systems \cite{Deligiannis2018,Nosrati2018}.

In MIMO radars, antenna selection has been studied mostly from the perspective of target parameter estimation. An optimal antenna placement was proposed in \cite{He2010} to minimize the Cram\'er-Rao lower bound (CRLB) of the velocity estimates. In \cite{Godrich2012} a combinatorial optimization approach was used to achieve resource allocation for localization error minimization in multiple radar systems. The CRLB for target location in MIMO radars with collocated antennas was derived in \cite{Gorji2014}, allowing its determinant to be minimized. Joint antenna subset selection and optimal power allocation were also implemented in \cite{Ma2014} for localization in MIMO radar sensor networks via convex optimization. In a similar vein, the idea of minimum redundancy has been successfully applied to the design of physical transmit/receive arrays to form MIMO virtual arrays with maximum contiguous aperture, i.e. minimum redundancy virtual arrays (MRVA) \cite{Chen2008b}. The two-level autocorrelation property of the difference sets (DSs) was then successfully exploited to maximize the virtual aperture \cite{JianDong2009}.

In this paper we address the problem of antenna selection for interference cancellation and SINR maximization. Antenna selection can be applied to the transmit and receive arrays separately, jointly to the transmit and receive arrays, or to the matched filter bank (virtual array).  We study all of these scenarios and propose a comprehensive optimization method to derive their solutions. We first examine the joint transmit/recieve element selection, which reduces the dimensionality and consequently decreases the computational cost. We then consider the factored selection approach in which we separately select subsets of the transmit and receive arrays. We formulate the factored problem as a coupled optimization such that both selections are solved together. The computational cost of the MIMO radar can also be alleviated by reducing the number of matched filters used to generate a virtual array at the receiver, which involves the application of element selection to the virtual array. Finally, we bring these scenarios together in a hybrid selection strategy that is capable of reducing the number of transmitters, receivers and  matched filters simultaneously in a unified approach.  

The main contributions of this paper are as follows.
\begin{enumerate}
    \item We express the output SINR, denoted as $\mt{SINR}_\mt{out}$, as a function of selected elements of the MIMO array in a scenario comprising a single target, multiple jammers, and clutter.
    \item We propose four different selection approaches, each achieving a different efficiency in terms of  hardware (e.g., baseband and RF), computational, and power cost. 
    \item Since the $\mt{SINR}_\mt{out}$ is a joint function of transmitters and receivers in MIMO, we propose a new factored problem formulation that permits us to decouple the transmit and receive sides and allows their designs to be performed separately. 
    \item We formulate the dual problem and study the performance of the proposed selection methods from a mathematical point of view.
    \item We propose a relaxation method and successfully approximate the global solution via a set of problem-specific randomized rounding strategies.
\end{enumerate}

The rest of this paper is organized as follows. In section~\ref{sec:problemformulation} we present the formulation of the $\mt{SINR}_\mt{out}$ maximization using element selection. We then study the selection approaches in Section \ref{sec:selection}. The relaxation strategy is detailed in Section \ref{sec:relaxation}, and the numerical results are presented in Section \ref{sec:sim}. Finally some conclusions are drawn in section \ref{sec:conclusion}.

\subsection*{Notation}
We use bold lower-case letters to denote vectors, and upper-case letters for matrices. The notation $\mathbb{E}$ is the expectation operator, and Tr($\bd M$) denotes the trace of $\bd M$. $(\bullet)^T$  and $(\bullet)^H$ are the Hermitian and transpose operations. The operation diag($\bd v$) constructs a square diagonal with $\bd v$ along the diagonal, whereas diag($\bd M$) extracts the diagonal of $\bd M$. The function real($\bullet$) takes the real part of its complex argument. We use $\otimes$ for Kronecker product. Finally, $\bd 1_N$ is a $N\times1$ vector of all ones, $\bd 0_N$ a vector of zeros, and $\bd I_N$ the $N\times N$ identity matrix.

\section{Problem Formulation}
\label{sec:problemformulation}
\begin{figure}[!t]
    \centering
    \includegraphics[clip, trim=3cm 9cm 15cm 26cm,width=9cm]{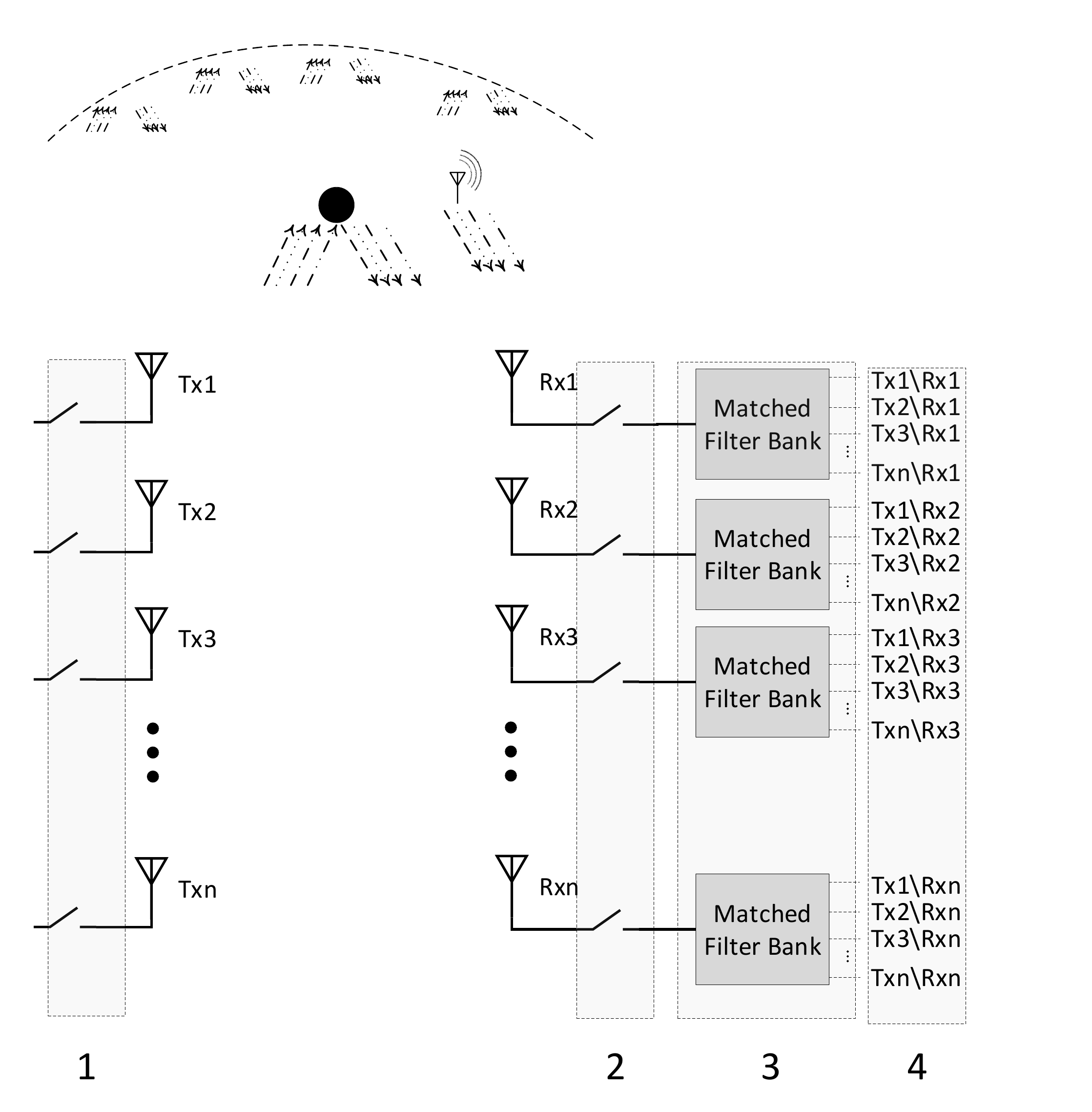}
    \caption{Block diagram of the considered MIMO system. Stage 1: transmit antenna selection. Stage 2: receive antenna selection. Stage 3: matched-filter selection in DSP. Stage 4: output signal selection.}
    \label{fig:MIMO_System}
\end{figure}

Let us consider a MIMO radar equipped with $M$ transmitters and $N$ receivers as shown in Fig. \ref{fig:MIMO_System}. Each transmitter emits one of the predesigned orthogonal waveforms from the waveform vector $\boldsymbol{\phi}(t)=\left[ \phi_1(t),\phi_2(t),...,\phi_M(t) \right]$. The snapshot vector received by the receive array for pulse $\tau$ is
\begin{align}
\nonumber\bd x(t,\tau)=\bd x_\mt s(t,\tau)+\bd x_\mt c(t,\tau)+\bd x_\mt j(t,\tau)+\bd x_\mt n(t,\tau),
\end{align}
where $\bd x_c$ and $\bd x_j$ represent the clutter and jammer respectively. The signal of interest (SOI), $\bd x_s$, represents the target reflection and $\bd x_{n}$ is zero-mean additive Gaussian noise with variance $\sigma_n^2$.
By applying matched-filtering with respect to the $M$ orthogonal waveforms, the extended receive signal becomes
\begin{align}
\nonumber \tilde{\bs x}(\tau)&=\mt{vec}\left(\int_{T_p} \bd x(t,\tau)\bs \phi^H(t) dt\right)\\
\nonumber&=\tilde{\bs x}_{\mt{s}}(\tau)+\tilde{\bs x}_{\mt{c}}(\tau)+\tilde{\bs x}_{\mt{j}}(\tau)+\tilde{\bs x}_{\mt{n}}(\tau),
\end{align}
where $\tilde{\bs x}$ represents the vectorized version of the matched-filtered snapshot vector. Exploiting the orthogonality assumption, we can write the SOI as
\begin{align}
\nonumber \tilde{\bs x}_{\mt{s}}(\tau)&= \beta_s\bd a_{\mt{s}}\\
\nonumber\bd a_{\mt{s}}&=\bd a_{\mt t}(\theta_{\mt t}) \otimes \bd a_{\mt r}(\theta_{\mt r}),
\end{align}
where $\beta_\mt s$ is the target reflection coefficient, which we assume obeys the  Swirling \Rmnum{2} model. The transmit steering vector, $\bd a_{\mt t}(\theta_\mt{t})$, corresponds the Direction-of-Departure (DoD) $\theta_\mt t$, and the receive steering vector, $\bd a_{\mt r}(\theta_{\mt r})$, is associated with DoA $\theta_\mt r$. In the case of a ULA, the steering vectors are given by 
 \begin{align}
	\nonumber\bd a_{\mt t}(\theta_\mt t)&=[1,e^{j 2\pi d_\mt t sin\theta_\mt t},e^{j 2\pi 2 d_\mt t sin\theta_\mt t},...,e^{j 2\pi M d_\mt t sin\theta_\mt t}],\\
	\nonumber\bd a_{\mt r}(\theta_\mt r)&=[1,e^{j 2\pi d_\mt r sin\theta_\mt r},e^{j 2\pi 2 d_\mt r sin\theta_\mt r},...,e^{j 2\pi N d_\mt r sin\theta_\mt r}],
\end{align}
with $d_\mt t,d_\mt r$ denoting the inter-element spacing employed in transmit and receive arrays.

Now assuming a set of angle cells, $\left\lbrace\theta_i\right\rbrace_{i=1}^{N_\mt c}$, we model the clutter as the reflections from these directions and extract the received clutter signal as
\begin{align}
\nonumber\tilde{\bs x}_{\mt{c}}(\tau)&=\sum\limits_{i=1}^{N_{\mt c}}\beta_{i}\bd a_{\mt{c},i}\\
\nonumber\bd a_{\mt{c},i}&=\bd a_{\mt t}(\theta_{\mt {tc},i}) \otimes \bd a_{\mt{r}}(\theta_{\mt{ rc},i}),
\end{align}
where $\beta_i$ is the reflection coefficient of the $i$-th clutter cell from directions $\theta_{\mt {tc},i}$, and $\theta_{\mt {rc},i}$ with respect to transmit, and receive sides. Suppose that $N_\mt j$ jamming signals are in the field of view of the radar. Then the jamming signal is expressed as
\begin{align}
\tilde{\bs x}_{\mt{j}}(\tau)&=\sum\limits_{i=1}^{N_{\mt j}}\alpha_{i}\bd a_{\mt{j},i}\\
\bd a_{\mt{j},i}&=\tilde{\bs x}_{\mt{j,i}}(\tau) \otimes \bd a_{\mt r}(\theta_{\mt{rj},i}),
\end{align}
with $\alpha_{i}$, and $\tilde{\bs x}_{\mt{j,i}}(\tau)$ denoting the complex amplitude, and the matched filtered version of the $i$-th jamming signal. Given a strong jamming source $ \bd x_{\mt{j,i}}(t,\tau)$ with a power of $\hat{\alpha_{i}}$ , which emulates the radar orthogonal waveforms we have
\begin{align}\label{eq:jamm_assumption}
 \bd x_{\mt{j,i}}(t,\tau)=\hat{\alpha_{i}}\sum\limits_{i=1}^{M}\phi_i(t,\tau),
\end{align}
then, this signal passes through the matched filters and by incorporating $\hat{\alpha_{i}}$ in $\alpha_{i}$ we can write $\tilde{\bs x}_{\mt{j,i}}(\tau)=\bd 1$ \cite{Li2014,Vaidyanathan2009}. The received signal is then input to an adaptive filter with weights vector, $\bd w$, giving the output 
\begin{align}
y(\tau)=\bd w^H \tilde{\bs x}(\tau).
\end{align}
The weights vector that preserves the SOI, $\tilde{\bs x}_{\mt{s}}$, while suppressing the clutter, jammers and noise, thus  maximizing the output SINR, is obtained by solving the following optimization
\begin{align}
\min_{\bd w}\; &\bd w^H \bd R \bd w\\
\textrm{s.t.}\;\;& \bd w^H\bd a_{\mt{s}}=1.
\end{align}
This yields the solution \cite{VanTrees2002}
\begin{align}\label{eq:weight}
\bd w=\frac{\bd R^{-1}\bd a_{\mt{s}} }{\bd a_{\mt{s}}^H\bd R^{-1}\bd a_{\mt{s}}},
\end{align}
where $\bd R$ is the interference (jamming and clutter) plus noise covariance matrix of size $MN\times MN$. Assuming the clutter, jammers and noise are statistically independent, we have
\begin{align}
\nonumber\bd R&=\bd R_\mt c+\bd R_\mt j+\bd R_n.
\end{align}
Taking the clutter scattering coefficients, $\beta_i$ to be mutually uncorrelated, we find that 
\begin{align}
    \nonumber\bd R_\mt c&=\mathbb{E}\lbrace\tilde{\bs x}_{\mt{c}}(\tau)\tilde{\bs x}^H_{\mt{c}}(\tau)\rbrace=\sum\limits_{i=1}^{N_{\mt{c}}}\sigma^2_{\mt{c},i}\bd a_{\mt{c},i}\bd a_{\mt{c},i}^H=\bd A_\mt c \bd P_\mt c \bd A_\mt c^H,
\end{align}
where
\begin{align}
\sigma^2_{\mt{c},i}&=\mathbb{E}\lbrace\beta_{i}\beta_{i}^*\rbrace,\\
\bd A_\mt c&=[a_{\mt{c},1},a_{\mt{c},2},...,a_{\mt{c},N_\mt c}],\;\; MN\times N_\mt c
\end{align}
and 
\begin{align}
  \bd P_\mt c=\mt{diag}\left(\sigma^2_{c,1},..., \sigma^2_{c,N_\mt c}\right).
\end{align}

Similarly, using (\ref{eq:jamm_assumption}), the covariance matrix for the jamming signal is found to be
\begin{align}
\nonumber\bd R_\mt j=\mathbb{E}\lbrace\tilde{\bs x}_{\mt{j}}(\tau)\tilde{\bs x}^H_{\mt{j}}(\tau)\rbrace=\bd A_\mt j \bd P_\mt j \bd A_\mt j^H,
\end{align}
where $\bd A_\mt j$, and $\bd P_\mt j$ are $MN\times MN_\mt j$, and $ MN_\mt j \times MN_\mt j$  matrices denoted as
\begin{align}
    \bd A_\mt j&=[\underbrace{\bd a_{\mt r}(\theta_{{\mt{rj},1}}),...,\bd a_{\mt r}(\theta_{{\mt{rj},1}})}_{M\;\mt{times}},...,\underbrace{\bd a_{\mt r}(\theta_{{\mt{rj},N_\mt j}}),...,\bd a_{\mt r}(\theta_{{\mt{rj},N_\mt j}})}_{M\;\mt{times}}],\\
     \bd P_\mt j&=\mt{diag}(\underbrace{\sigma^2_{j,1},...,\sigma^2_{j,1}}_{M\;\mt{times}},...,\underbrace{\sigma^2_{j,N_\mt j},...,\sigma^2_{j,N_\mt j}}_{M\;\mt{times}}),
\end{align}
and
\begin{align}
\sigma^2_{\mt{j},i}=\mathbb{E}\lbrace\alpha_{i}\alpha_{i}^*\rbrace.
\end{align}
Finally, the covariance matrix of the white noise is given by
\begin{align}
    \nonumber\bd R_n&=\mathbb{E}\lbrace\tilde{\bs x}_{\mt{n}}(\tau)\tilde{\bs x}^H_{\mt{n}}(\tau)\rbrace=\sigma_n^2\bd I_{MN}.
\end{align}
Now writing $\sigma^2_{\mt{s}}=\mathbb{E}\lbrace\beta_{\mt{s}}\beta_{\mt{s}}^*\rbrace$, the SINR at the output of the filter (\ref{eq:weight}) is
\begin{align}\label{eq:sinr_out}
\mt{SINR}_{\mt{out}}=\sigma^2_{\mt{s}}\bd a_{\mt{s}}^H\bd R^{-1}\bd a_{\mt{s}}.
\end{align}
The inverse covariance matrix becomes
\begin{align}
    \bd R^{-1}&=\left(\bd A_\mt c \bd P_\mt c \bd A_\mt c^H+\bd A_\mt j \bd P_\mt j \bd A_\mt j^H+\sigma_n^2\bd I_{MN}\right)^{-1}\\
    &\stackrel{\text{(a)}}{=}\left(\bd A_\mt{jc} \bd P_\mt{jc} \bd A_\mt{jc}^H+\sigma_n^2\bd I_{MN}\right)^{-1}\\
    &\stackrel{\text{(b)}}{=}\sigma^{-2}_n \left( I_{MN}-\bd A_{\mt{jc}}\left(\sigma^2_n \bd P_\mt{\mt{jc}}^{-1}+\bd A_{\mt{jc}}^H\bd A_{\mt{jc}} \right)^{-1}\bd A_{\mt{jc}}^H \right),
\end{align}
where in (a) we put $\bd A_{\mt{jc}}=[\bd A_\mt c,\bd A_\mt j] $, and $\bd P_{\mt{jc}}=\mt{diag}\left(\bd P_\mt c,\bd P_\mt j\right) $, and (b) follows from the matrix inversion lemma. Now, we can reformulate the output SINR as
\begin{align}
    \mt{SINR}_{\mt{out}}&=\frac{\sigma^2_{s}}{\sigma^2_{n}}\bd a_s^H\big(\bd I_{MN}-\bd A_{\mt{jc}}(\sigma^2_n \bd P_{\mt{jc}}^{-1}+\bd A_{\mt{jc}}^H\bd A_{\mt{jc}} )^{-1}\bd A_{\mt{jc}}^H \big)\bd a_s.
\end{align}
Defining  the matrices
\begin{align}
    \bd B_{\mt{jc}}=\sigma_n^2\bd P_{\mt{jc}}^{-1},\;\;\bd A_{\mt{s}}=[\bd a_s,\bd A_{\mt{jc}}],\;\;
    \bd B_{\mt{s}}=\left[\bd 0,\sigma_n^2\bd P_{\mt{jc}}^{-1}\right],
\end{align}
and making use of the determinant formula of block matrices, we obtain\cite{Wang2016,Deligiannis2018}
\begin{align}
\nonumber\left|\bd A_{\mt{s}}^H  \bd A_{\mt{s}}+\bd B_{\mt{s}}\right|&=\left|\begin{matrix}
    \bd a_s^H \bd a_s       & \bd a_s^H \bd A_\mt{jc}  \\
    \bd A_\mt{jc}^H \bd a_s      & \bd A_{\mt{jc}}^H  \bd A_{\mt{jc}}+\bd B_{\mt{jc}}  \\
\end{matrix}\right|\\
\nonumber&=\left|\bd A_{\mt{jc}}^H  \bd A_{\mt{jc}}+\bd B_{\mt{jc}}\right|\\
\times&\bd a_s^H\left(\bd I_{MN}-\bd A_\mt{jc}\left(\bd B_{\mt{jc}}+\bd A_\mt{jc}^H\bd A_\mt{jc} \right)^{-1}\bd A_\mt{jc}^H \right)\bd a_s.
\end{align}
Substituting this into the expression of $\mt{SINR}_{out}$ yields
\begin{align}
   \nonumber \mt{SINR}_{\mt{out}}&=\frac{\sigma^2_{s}}{\sigma^2_{n}}\frac{\left|\bd A_{\mt{s}}^H  \bd A_{\mt{s}}+\bd B_{\mt{s}}\right|}{\left|\bd A_{\mt{jc}}^H  \bd A_{\mt{jc}}+\bd B_{\mt{jc}}\right|},
\end{align}
where $\left|.\right|$ denotes the determinant and the signal-to-noise ratio, $\mt{SNR}= \frac{\sigma^2_{s}}{\sigma^2_{n}}$. This reveals that, although the SNR is constant, the set of active transmit and receive elements directly affects the achieved output SINR by varying the interplay among the jamming and clutter steering vectors and their powers. Let us introduce a binary vector $\bd c$ with elements 0 if their corresponding elements are inactive and 1 otherwise. Then the element selection can be incorporated into the expression of the output $\mt{SINR}$ as follows
\begin{align}
\mt{SINR}_{\mt{out}}(\bd c)=\frac{\sigma^2_{s}}{\sigma^2_{n}}h(\bd c),
\end{align}
where
\begin{align}
    h(\bd c)=\frac{\left|\bd A_{\mt{s}}^H \mt{diag}(\bd c) \bd A_{\mt{s}}+\bd B_{\mt{s}}\right|}{\left|\bd A_{\mt{jc}}^H \mt{diag}(\bd c)  \bd A_{\mt{jc}}+\bd B_{\mt{jc}}\right|}.
\end{align}
The optimum set of $k$ active elements that maximizes the SINR is found via the following optimization:
\begin{align}\label{eq:maxdet_sinr}
	\max_{\bd c}\,\,& \mt{SINR}_{\mt{out}}(\bd c)\\
	\mathrm{s.t.}\,\,&  c_{i}^2-c_i=0\,\,\,\,\, i=1...MN,\\
	&\bd c^T\bd c=k.
\end{align}
The optimization in (\ref{eq:maxdet_sinr}) is maximization of the volume of two ellipsoids. Therefore, we may employ  log-determinant function as
\begin{align}\label{eq:maxdet_ref}
    \nonumber f(\bd c)=&\mt{log det} \left(h\left(\bd c\right)\right)\\
    \nonumber=&\mt{log det}\left(\bd A_{\mt{s}}^H \mt{diag}(\bd c) \bd A_{\mt{s}}+\bd B_{\mt{s}}\right)\\
    &-\mt{log det}\left(\bd A_{\mt{jc}}^H \mt{diag}(\bd c)  \bd A_{\mt{jc}}+\bd B_{\mt{jc}}\right).
\end{align}

This problem can be effectively solved via a log-determinant relaxation and a sequential convex programming (SCP) procedure accordingly. We will elaborate on the solution approximation in Section \ref{sec:relaxation}.

Considering (\ref{eq:maxdet_sinr}), we propose four different methods to apply selection in a MIMO radar. We list all the requirements in different modes in a quadratic form. Hence, we cast the general problem of antenna selection in a MIMO radar as follows
\begin{align}\label{eq:opt_pr_1}
	\max_{\bd c}\,\,&f(\bd c)\\
	\,\,& \bd c^{T}\bd W_i\bd c+\bd q_i^T\bd c+r_i\leq 0,\;\;\;\;i=1,...,\ell.
\end{align}
where $\bd W_i \in  \mathbb{S}^{MN}$, $\bd q_i \in  \mathbb{R}^{MN}$, and $r_i \in \mathbb{R}$. Note that $\mathbb{S}^n$ denotes the set of $n\times n$ symmetric matrices, and $\mathbb{R}^n$ represents the set of real vectors of size $n$.

\section{Selection strategies} \label{sec:selection}
The selection strategy may be applied at each of the four stages of a MIMO radar depicted in Fig.~\ref{fig:MIMO_System}. In what follows, we detail and compare these selection approaches.
\subsection{Joint Tx-Rx selection}
    The first selection strategy involves thinning the MIMO virtual array by selecting the individual matched-filters at the output (stage 4) of Fig.~\ref{fig:MIMO_System}. In this case, we reshape the selection vector $\bd c$ as a matrix $\bd C$
%
\begin{align}\label{eq:sel_mat}
\bd C = \bordermatrix{~ & 		\text{Tx}_1 	& \text{Tx}_2 &\cdots		&\text{Tx}_M \cr
                  \text{Rx}_1 & 	c_{1,1} 				& 	c_{1,2}					& \cdots	&  c_{1,M} 	\cr
                  \text{Rx}_2 & 	c_{2,1} 				& 	c_{2,2}					& \cdots	&  c_{2,M}	\cr
									\vdots 			& \vdots  		& 	\vdots 		& \ddots	&  \vdots 	\cr
									\text{Rx}_N & 	c_{N,1}  				& 	c_{N,2}					& \cdots	&  c_{N,M}	\cr},
\end{align}
such that $\bd c=\mt{vec}(\bd C)$, and $c_{i,j}$ is the entry that indicates whether the $j$-th matched filter (which extracts the $j$-th waveform) in the $i$-th receiver is selected. The joint thinning mode is obtained by directly selecting $k$ elements in $\bd c$. The selection problem of (\ref{eq:opt_pr_1}) can now be written as
\begin{align}\label{eq:maxdet_f}
	\nonumber\max_{\bd c}\,\,& f(\bd c)\\
	\nonumber\mathrm{s.t.}\,\,&  c_{i}^2-c_i=0\,\,\,\,\, i=1...MN,\\
	&\bd c^T\bd c=k.
\end{align}
%
In this formulation, we place a constraint only on the number of output signals and any subset of the output matched filters is a possible solution. The joint selection finds the best subset, which decreases the dimensionality of the signal used in the post processing (e.g. for detection, estimation or other tasks). However, the entire system including transmitters, receivers, and matched filters should be active, and consequently the hardware cost and power consumption remain high.      

\subsection{Factored Tx and Rx selection}\label{subsec:factored}
As all of the transmitters are required to be active in the joint selection strategy, transmitters that do not contribute to the selected set of matched-filters effectively waste the power alloted to them. This issue can be mitigated by factorizing the selection problem into transmit and receive sub-problems. 
Suppose that we select $k_\mt t$ out of $M$ transmitting antennas (stage 1 in Fig.~\ref{fig:MIMO_System}) and $k_\mt r$ out of the available $N$ receive antennas (stage 2 in Fig.~\ref{fig:MIMO_System}). In terms of the selection matrix (\ref{eq:sel_mat}), this strategy selects $k_\mt r$ rows and $k_\mt t$ columns. Then the factored selection involves the  optimization of two selection vectors jointly, one for the transmitters and the other for the receivers. We now develop a novel way to reformulate this coupled problem in one unified formulation.  

Let $\bd V_{\mt t}=\{\bd v_{\mt t,1},...,\bd v_{\mt t,M}\}$, and $\bd V_{\mt r}=\{\bd v_{\mt r,1},...,\bd v_{\mt r,N}\}$ be a set of binary vectors each of which denoting a specific transmitter or receiver in the selection matrix 
\begin{align}
   \bd v_{\mt t,i} &= \mt{vec}\left(\left[\bd 0,\hdots,\underset{\text{Tx}_i  }{\bd 1},\hdots,\bd 0\right] \right)\\
   \bd v_{\mt r,i} &= \mt{vec}\left(\left[\bd 0,\hdots,\underset{\text{Rx}_i  }{\bd 1},\hdots,\bd 0\right]^T \right)\\
\end{align}

Then, the factored selection problem may be expressed as
\begin{subequations}\label{eq:factored_new_form}
\begin{align+}
	\nonumber\max_{\bd c}\,\,& f(\bd c)\\
	\label{subeq:bin}\mathrm{s.t.}\,\,&  c_{i}^2-c_i=0\,\,\,\,\, i=1...MN,\\
	\label{subeq:TxMem}\,\,& \bd c^{T}\bd P_{{\mt t,i}}\bd c\in \{0,k_\mt r\}\,\,\,\,\, i=1...M,\\
	\label{subeq:RxMem}\,\,& \bd c^{T}\bd P_{{\mt r,i}}\bd c\in \{0,k_\mt t\}\,\,\,\,\, i=1...N,\\
\label{subeq:TotNum}\,& \bd c^{T}\bd c=k_\mt t k_\mt r,
	\end{align+}	
	\end{subequations}
where
\begin{align}
\label{eq:mem_mat_Tx}\bd P_{{\mt t,i}}=\text{diag}(\bd v_{\mt t,i}), i=1,...,M,\;\bd P_{{\mt r,i}}=\text{diag}(\bd v_{\mt r,i}), i=1,...,N.
	\end{align}
In (\ref{subeq:TxMem}) and (\ref{subeq:RxMem}) we constrain the  number of active elements in each column (row) to be exactly 0 or $k_\mt t$ (0 or $k_\mt r$).

Now let us define $\bd Q$ as the rectangular matrix 
\begin{align}
\label{eq:Q}\bd Q&=\bd P\bd P^T,\text{ where}\;\;\bd P=\left[\bd v_{\mt t,1},...,\bd v_{\mt t,M},\bd v_{\mt r,1},...,\bd v_{\mt r,N}\right].
\end{align}

\begin{theorem}\label{theo:added_constraints}
Let $\mathcal S$ be the set of selection vectors in conjunction with a factored selection problem comprising of $k_\mt t$ and $k_\mt r$ out of $M$ transmitters and $N$ receivers respectively. Then $\mathcal S$ is given by
\begin{align}
	\nonumber\,\,& \mathcal S=\mathcal S_1 \cap \mathcal S_2 \cap \mathcal S_3\cap \mathcal S_4,
	\end{align}
where the sets $\mathcal S_1$ - $\mathcal S_4$ are defined as
\begin{align}
	\nonumber\,\,& \mathcal S_1=\{ \bd c\;|\; \bd c^{T}\bd Q\bd c=k_\mt rk_\mt t (k_\mt r+k_\mt t)\}\\
	\nonumber\,\,& \mathcal S_2=\{ \bd c\;|\;\bd c^{T}\bd P_{{\mt t,i}}\bd c\leq k_\mt r\,\,\,\,\, i=1...M,\}\\
	\nonumber\,\,& \mathcal S_3=\{ \bd c\;|\;\bd c^{T}\bd P_{{\mt r,i}}\bd c\leq k_\mt t\,\,\,\,\, i=1...N\}\\
	\nonumber\,\,& \mathcal S_4=\{ \bd c\;|\;\bd c^{T}\bd c= k_\mt t k_\mt r\}.
	\end{align}

\end{theorem}
\textbf{Proof:} See Appendix~\ref{sec:appendix_Proof_theorem1}.

Like the constraint in (\ref{subeq:bin}), the binary constraints involving the quadratic forms in (\ref{subeq:TxMem}) and (\ref{subeq:RxMem}) are non-convex. Therefore, we propose relaxing them by employing the following set of quadratic constraints instead 
%
\begin{align}
 \label{sueq:first}\bd c^{T}\bd Q\bd c&=k_\mt rk_\mt t (k_\mt r+k_\mt t)\\
	 \label{sueq:second}\bd c^{T}\bd P_{{\mt t,i}}\bd c&\leq k_\mt r\,\,\,\,\, i=1...M\\
	\label{sueq:third} \bd c^{T}\bd P_{{\mt r,i}}\bd c&\leq k_\mt t\,\,\,\,\, i=1...N,
	\end{align}
Using Theorem~\ref{theo:added_constraints}, the factored selection problem becomes
\begin{subequations}\label{eq:fct}
\begin{align+}
	\nonumber\max_{\bd c}\,\,& f(\bd c)\\
	 \mathrm{s.t.}\;\;&  c_{i}(c_i-1)=0\,\,\,\,\, i=1...MN,\label{eq:fct_bin}\\
	&\bd c^{T}\bd c=k_\mt tk_\mt r \label{eq:fct_k}\\
	& \bd c^{T}\bd Q\bd c= k_\mt rk_\mt t (k_\mt r+k_\mt t )\label{eq:fct_fct}\\
	& \bd c^{T}\bd P_{{\mt t,i}}\bd c\leq k_\mt r \,\,\,\,\, i=1...M,\\
	& \bd c^{T}\bd P_{{\mt r,i}}\bd c\leq k_\mt t \,\,\,\,\, i=1...N.
	\end{align+}
    \end{subequations}

We recast the added binary constraints in the factored problem, into a quadratic form as a special case of (\ref{eq:opt_pr_1}). This enables us to compare the performance of the factored selection with that of the joint selection. We show that the optimum solution (i.e. SINR) obtained by the Lagrange dual of the joint selection optimization is always greater than or equal to that yielded by the factored problem.   
To this end, we derive a dual problem for the factored selection problem in (\ref{eq:fct}).
We revise the factored problem (\ref{eq:fct}), by introducing new variables $\bd X$, and $\bd Y$ as 
\begin{subequations}\label{eq:fact_min_equi}
\begin{align+}
	\nonumber\min_{\bd c}\,\,& \mt{log det}(\bd X^{-1})-\mt{log det}(\bd Y^{-1})\\
	 \nonumber\mathrm{s.t.}&\\
	 \bs \Lambda &: \bd X=\bd A_\mt{s}^H \mt{diag}(\bd c) \bd A_\mt{s}+\bd B_\mt{s}\\
	 \bs \Delta &: \bd Y=\bd A_\mt{jc}^H \mt{diag}(\bd c) \bd A_\mt{jc}+\bd B_\mt{jc}\\
	 \mu_i&: \;\; c_{i}(c_i-1)=0\,\,\,\,\, i=1...MN,\\
		 \nu&:\;\; \bd c^{T}\bd c=k_\mt tk_\mt r\\
 \lambda&:\;\; \bd c^{T}\bd Q\bd c\leq k_\mt rk_\mt t (k_\mt r+k_\mt t)\\
\rho_i&: \bd c^{T}\bd P_{{\mt t,i}}\bd c\leq k_\mt r\,\,\,\,\, i=1...M,\\
		\eta_i&: \bd c^{T}\bd P_{{\mt r,i}}\bd c\leq k_\mt t \,\,\,\,\, i=1...N,
	\end{align+}
    \end{subequations}
with Lagrange multipliers $\bs \Lambda\in \mathbb S^{n+1} (n=N_\mt c+N_\mt j)$, $\bs \Delta \in \mathbb S^{n}$, $\bs \mu \in \mathbb R^{MN}$, $\bs \rho \in \mathbb R^M$, $\bs \eta \in \mathbb R^N$, and $\nu,\lambda\in \mathbb R$. We then introduce the Lagrangian
\begin{align}
     \nonumber L^{\mt{fct}}(&\bd c,\bd X,\bd Y,\bs \Lambda,\bs \Delta,\bs \mu, \nu,\lambda,\bs \rho,\bs \eta)=\\
   \nonumber &\mt{log det}(\bd X^{-1})-\mt{log det}(\bd Y^{-1})+\mt{tr}(\bd X\bs \Lambda)+\mt{tr}(\bd Y\bs\Delta)\\
  \nonumber  &+\bd c^T(\text{diag}( \bs\mu)+\nu\bd I
 +\lambda \bd Q+\sum\limits_{i=1}^{M}\rho_i\bd P_{t,i}+\sum\limits_{i=1}^{N}\eta_i\bd P_{r,i} )\bd c\\
    \nonumber &-\bs \Lambda\left(\bd A_\mt{s}^H \mt{diag}(\bd c) \bd A_\mt{s}+\bd B_\mt{s}\right)-\bs \Delta\left(\bd A_\mt{jc}^H \mt{diag}(\bd c) \bd A_\mt{jc}+\bd B_\mt{jc}\right)\\
    \nonumber &-\bs \mu^T\bd c-k_\mt tk_\mt r\nu-\lambda k_\mt rk_\mt t (k_\mt r+k_\mt t)\\
&-k_\mt r\sum\limits_{i=1}^{M}\rho_i-k_\mt t\sum\limits_{i=1}^{N}\eta_i  .
\end{align}
By rearranging the Lagrangian we get
\begin{align}
     \nonumber L^{\mt{fct}}(&\bd c,\bd X,\bd Y,\bs \Lambda,\bs \Delta,\bs \mu, \nu,\lambda,\bs \rho,\bs \eta)=\\
   \nonumber &\mt{log det}(\bd X^{-1})-\mt{log det}(\bd Y^{-1})+\mt{tr}(\bd X\bs \Lambda)+\mt{tr}(\bd Y\bs\Delta)\\
   \nonumber& -\mt{tr}(\bd B_\mt{s}\bs\Lambda)-\mt{tr}(\bd B_\mt{jc}\bs\Delta)\\
   +\nonumber & \sum\limits_{i=1}^{MN} c_i \underbrace{\left(\left[\bd A_{\mt{s}}^T\right]_i^T\bs \Lambda \left[\bd A_{\mt{s}}^H\right]_i +\left[\bd A_{\mt{jc}}^T\right]_i^T\bs \Delta \left[\bd A_{\mt{jc}}^H\right]_i +\mu_i\right)}_{\omega_i}\\
  \nonumber  &+\bd c^T(\text{diag}( \bs\mu)+\nu\bd I
 +\lambda \bd Q+\sum\limits_{i=1}^{M}\rho_i\bd P_{t,i}+\sum\limits_{i=1}^{N}\eta_i\bd P_{r,i} )\bd c\\
    \nonumber &-k_\mt tk_\mt r\nu-\lambda k_\mt rk_\mt t (k_\mt r+k_\mt t)\\
&-N\sum\limits_{i=1}^{M}\rho_i+M\sum\limits_{i=1}^{N}\eta_i  .
\end{align}
We minimize $L^{\mt{fct}}$ with respect to $\bd c$, $\bd X$, and $\bd Y$. Noting that $L^{\mt{fct}}$ is a mixture of two volume covering ellipsoids in terms of $\bd X$, and $\bd Y$ (see p 222 in \cite{Boyd2010b}, and Appendix in \cite{Joshi2009}) and given the  set of quadratic forms in $\bd c$, we arrive at the Lagrange dual function $g^{\mt{fct}}$ in (\ref{eq:lgrngdufun_fac}). 
\begin{figure*}[!t]
\normalsize
\setcounter{MYtempeqncnt}{\value{equation}+1}
\begin{align}
\nonumber &g^{\mt{fct}}(\bs \Lambda,\bs \Delta, \bs \mu, \nu,\lambda,\bs \rho,\bs \eta)=\inf_{\bd c,\bd X,\bd Y} L^{\mt{fct}}(\bd c,\bd X,\bd Y,\bs \Lambda,\bs \Delta,\bs \mu, \nu,\lambda,\bs \rho,\bs \eta)\}\\
  &=\left\{ 
                \begin{array}{llll}
                 \mt{log det}(\bs \Lambda)+ \mt{log det}(\bs \Delta)+2n+1-\mt{tr}(\bs \Lambda \bd B_\mt{s})-\mt{tr}(\bs \Delta \bd B_\mt{jc}) - \frac{1}{4} \boldsymbol \omega^T\bigg(\text{diag}( \bs\mu)+\nu\bd I+\lambda \bd Q+\sum\limits_{i=1}^{M}\rho_i\bd P_{t,i}
									+\sum\limits_{i=1}^{N}\eta_i\bd P_{r,i} \bigg)^{-1}\boldsymbol \omega\\
								\hspace{9cm}	-k_\mt tk_\mt r\nu-\lambda k_\mt rk_\mt t (k_\mt r+k_\mt t)-k_\mt r\sum\limits_{i=1}^{M}\rho_i-k_\mt t\sum\limits_{i=1}^{N}\eta_i \\ 
					  	\label{eq:lgrngdufun_fac}\hspace{5cm}	\text{if}\;\; \bigg(\text{diag}( \bs\mu)+\nu\bd I+\lambda \bd Q+\sum\limits_{i=1}^{M}\rho_i\bd P_{t,i}
								+\sum\limits_{i=1}^{N}\eta_i\bd P_{r,i} \bigg)\succeq 0, \bd X\succeq 0,\bd Y\succeq 0 \\
									-\infty\;\;\;\;\;\;\; \text{otherwise.}
                \end{array}
								\right.
\end{align}
\setcounter{equation}{\value{MYtempeqncnt}}
\begin{align}
\nonumber &g^{\mt{jnt}}(\bs \Lambda,\bs \Delta, \bs \mu, \nu,\lambda,\bs \rho,\bs \eta)=\inf_{\bd c,\bd X,\bd Y} L^{\mt{jnt}}(\bd c,\bd X,\bd Y,\bs \Lambda,\bs \Delta,\bs \mu, \nu,\lambda,\bs \rho,\bs \eta)\}\\
  &=\left\{ 
                \begin{array}{llll}
                 \mt{log det}(\bs \Lambda)+ \mt{log det}(\bs \Delta)+2n+1-\mt{tr}(\bs \Lambda \bd B_\mt{s})-\mt{tr}(\bs \Delta \bd B_\mt{jc}) - \frac{1}{4} \boldsymbol \omega^T\bigg(\text{diag}( \bs\mu)+\nu\bd I+\lambda \bd Q+\sum\limits_{i=1}^{M}\rho_i\bd P_{t,i}
									+\sum\limits_{i=1}^{N}\eta_i\bd P_{r,i} \bigg)^{-1}\boldsymbol \omega\\
								\hspace{9cm}	-k^{\mt{jnt}}\nu-\lambda k^{\mt{jnt}} (M+N)-N\sum\limits_{i=1}^{M}\rho_i-M\sum\limits_{i=1}^{N}\eta_i \\ 
					  	\nonumber\label{eq:lgrngdufun_jnt}\tag{12}\hspace{5cm}	\text{if}\;\; \bigg(\text{diag}( \bs\mu)+\nu\bd I+\lambda \bd Q+\sum\limits_{i=1}^{M}\rho_i\bd P_{t,i}
								+\sum\limits_{i=1}^{N}\eta_i\bd P_{r,i} \bigg)\succeq 0, \bd X\succeq 0,\bd Y\succeq 0 \\
									-\infty\;\;\;\;\;\;\; \text{otherwise.}
                \end{array}
								\right.
\end{align}

\hrulefill
\vspace*{4pt}
\end{figure*}

\begin{theorem}\label{theo:fct}
Let $f^{\text{jnt}}$ be the optimal value associated with the joint selection (\ref{eq:maxdet_sinr}) and $f^{\text{fct}}$ be that of the factored selection (\ref{eq:fct}). Also, let $k_\mt r^{\mt{fct}}$, $k_\mt t^{\mt{fct}}$ be the number of selected transmitters and receivers in the factored selection, and $k^{\mt{jnt}}$ the number of jointly selected elements such that $k^{\mt{jnt}}=k_\mt r^{\mt{fct}}k_\mt t^{\mt{fct}}$. Then
\begin{equation}\label{eq:pr_1}
f^{\text{jnt}} \geq f^{\text{fct}}.
\end{equation}
\end{theorem}
%
\begin{IEEEproof}
Let us recast the joint problem as
\begin{subequations}\label{eq:jnt_ref}
\begin{align+}
	\nonumber\max_{\bd c}\,\,& f(\bd c)\\
	 	\nonumber\mathrm{s.t.}\\
	\mu_i&: \;\; c_{i}(c_i-1)=0\,\,\,\,\, i=1...MN,\\
		 \nu&:\;\; \bd c^{T}\bd c=k^{\text{jnt}}\\
	\label{eq:jnt_PP} \lambda&:\;\; \bd c^{T}\bd Q\bd c\leq k^{\text{jnt}} (M+N)\\
	\label{eq:jnt_Tx}\rho_i&: \bd c^{T}\bd P_{{\mt t,i}}\bd c\leq N\,\,\,\,\, i=1...M,\\
		\label{eq:jnt_Rx} \eta_i&: \bd c^{T}\bd P_{{\mt r,i}}\bd c\leq M \,\,\,\,\, i=1...N,
	\end{align+}
    \end{subequations}
We can also reformulate (\ref{eq:jnt_ref}) in terms of the equivalent minimization like (\ref{eq:fact_min_equi}) and derive the Lagrange dual function $g^{\mt{jnt}}$ as in (\ref{eq:lgrngdufun_jnt}). By minimization reformulation and employing $\bar{f}^{\text{jnt}}$, and $\bar{f}^{\text{fct}}$ as the corresponding optimal values, the expression in (\ref{eq:pr_1}) can be transformed into
\begin{equation}\label{eq:pr_2}
\bar{f}^{\text{jnt}} \leq \bar{f}^{\text{fct}}.
\end{equation}
Given the upper bounds in (\ref{eq:lgrngdufun_fac}) and (\ref{eq:lgrngdufun_jnt}), we  show (\ref{eq:pr_1}) by equivalently proving that 
\begin{align}
     g^{\mt{jnt}}\leq g^{\mt{fct}}.
\end{align}
%
%
Now we have that
\begin{align}
&\bd Q \succeq 0,\;\;\bd P_{t,i}\succeq 0, i=1,...,M&\;,\;\;
\bd P_{r,i}\succeq 0, i=1,...,N.
\end{align}
Also, the Karush-Kuhn-Tucker (KKT) conditions \cite{Boyd2010b} imply that 
\begin{align}
\lambda\geq 0,\;\;
\bs \rho \geq 0,\;\;
\bs \eta \geq 0.
\end{align}
Therefore, by subtracting (\ref{eq:lgrngdufun_fac}) and (\ref{eq:lgrngdufun_jnt}) we get
\begin{align}
\nonumber g^{\text{fct}}-g^{\text{jnt}}&=\lambda\Big(k^{\mt{jnt}} \big((M+N)-(k_\mt r^{\mt{fct}}+k_\mt t^{\mt{fct}})\big)\Big)\\
&\quad+\sum\limits_{i=1}^{M}\rho_i\left(N-k_\mt r^{\mt{fct}}\right)+\sum\limits_{i=1}^{N}\eta_i\left(M-k_\mt t^{\mt{fct}}\right)\\
\nonumber &\geq 0
\end{align}
\end{IEEEproof}
\setcounter{equation}{\value{equation}+1}

The factored selection operates on a subset of solutions that is included in the joint selection, and hence may not achieve the same optimal solution that is guaranteed by joint selection. Nonetheless, selecting a subset of transmitters allows the available total transmit power to be allocated only to the chosen elements. This is in contrast to the joint selection problem where all transmitters must be operational to guarantee that all matched filters are available for selection. Thus, assuming a total available transmit power $P_e = P_T$, the transmit power per element in the factored case is $P_e=\frac{P_t}{k_\mt t}$ as opposed to $\frac{P_t}{M}$ for the joint selection case.
It is important to note, however that increasing the allocation of transmit power per element may be restricted by the hardware limitations of the components in the RF chain, such as amplifier linear range. This may limit the gain achievable by the factored approach.
\subsection{Matched Filter Constrained Selection}\label{subsec:mfc}
We can adjust the transmitter power and SNR by a factored selection. Moreover, the number of receivers is decreased, which leads to a considerable hardware reduction. 
Since the number of transmitters is reduced in a factored selection, the spatial diversity is reduced significantly\cite{Nosrati2017c}. 
To preserve the spatial diversity provided by MIMO arrays but still reduce hardware and computation overheads, we propose restricting the number of matched filters in each receiver, as well as the number of receivers, in a matched filter constrained (MFC) selection strategy \cite{Nosrati2017}. Using this, we decrease the number of RF front-ends on the receive end (stage 2 in Fig.\ref{fig:MIMO_System}) as well as the required processing blocks in DSP (stage 3 in Fig.\ref{fig:MIMO_System}).   

We specify the MFC selection to select $k_\mt m$ matched filters in $k_\mt r$ receivers as follows
\begin{subequations}\label{eq:mfc}
\begin{align+}
	\nonumber\max_{\bd c}\,\,& f(c)\\
	\mathrm{s.t.}\,\,& c_{i}^2-c_i=0\,\,\,\,\, i=1...MN,\\
	&\bd c^{T}\bd c=k_\mt mk_\mt r,\\
	&\bd c^{T}\bd Q_\mt r\bd c=k_\mt m^2 k_\mt r,\\
	& \bd c^{T}\bd P_{{\mt t,i}}\bd c<=k_\mt r\,\,\,\,\, i=1...M,\\
	& \bd c^{T}\bd P_{{\mt r,i}}\bd c<=k_\mt m\,\,\,\,\, i=1...N,
	\end{align+}
	\end{subequations}
where $\bd Q_\mt r$ is
\begin{align}
\bd Q_\mt r=\bd P_\mt r\bd P_\mt r^T,\;\;\bd P_\mt r=[\bd v_{\mt r,1},...,\bd v_{\mt r,N}].
\end{align}
Comparing (\ref{eq:mfc}) with (\ref{eq:fct}) makes it obvious that the factored selection is a special case of the MFC selection. Therefore, for concision we state the following theorem without proof.
\begin{theorem}\label{theo:mfc_joint}
Let $f^{\text{mfc}}$ be the optimal value of the MFC selection (\ref{eq:mfc}) and $f^{\text{jnt}}$ the optimal value of the joint selection (\ref{eq:jnt_ref}), and $f^{\text{fct}}$ the optimal value of the factored selection (\ref{eq:fct}). Moreover, let $k_\mt m^{\mt{mfc}}$ be the number of selected matched-filters and receivers in MFC selection , $k_\mt r^{\mt{mfc}}$, let $k_\mt r^{\mt{fct}}$, $k_\mt t^{\mt{fct}}$ be the number of selected transmitters and receivers respectively, in factored selection, and $k^{\mt{jnt}}$ be the number of jointly selected elements such that $k^{\mt{jnt}}=k_\mt m^{\mt{mfc}}k_\mt r^{\mt{mfc}}=k_\mt r^{\mt{fct}}k_\mt t^{\mt{fct}}$. Then
\begin{equation}
 f^{\text{jnt}}\geq f^{\text{mfc}}\geq f^{\text{fct}}  .
\end{equation}
\end{theorem}
\subsection{Hybrid Selection}\label{subsec:hybrid}
By the giving up of spatial diversity described for factored selection , there is a benefit for transmit power and  selecting the optimum subset of matched filters with a decrease in the number of transmitters and receivers. On the other hand, in MFC selection we maintain spatial diversity at the expense of transmit power.
In this subsection, we propose integrating these two methods into a hybrid algorithm by which the MIMO array is entirely controlled. In a MIMO radar comprising $M$ transmitters and $N$ receivers, the hybrid selection finds the optimum subset including $k=k_\mt r k_\mt m$ matched filters such that exactly $k_\mt m$ out of $k_\mt t$ active transmitters are used in each $k_\mt r$ receivers.  By adopting the same methodology introduced in (\ref{eq:factored_new_form}), we define the hybrid selection as
    \begin{subequations}\label{eq:hybrid}
\begin{align+}
	\nonumber\max_{\bd c}\,\,& f(\bd c)\\
 \label{subeq:hybrid:bin}	\mathrm{s.t.}\,\,& c_{i}^2-c_i=0\,\,\,\,\, i=1...MN,\\ \label{subeq:hybrid:TotNum}\,& \bd c^{T}\bd c=k_\mt mk_\mt r,\\
 \label{subeq:hybrid:TxMem}\,\,& \bd c^{T}\bd P_{{\mt t,i}}\bd c\in \{0,k_\mt r\}\,\,\,\,\, i=1...M,\\
    \label{subeq:hybrid:RxMem}\,\,& \text{card}\left(\bd c^T \bd P_r \right)\leq k_\mt t,
    \end{align+}    
    \end{subequations}
where $\text{card}(\bd x)$ denotes the cardinality, the number of non-zero elements of vector $\bd x$. We first relax (\ref{subeq:hybrid:TxMem}) via (\ref{subeq:hybrid:relax1:TxMem}), and (\ref{subeq::hybrid:relax1:addconst1}) in the following optimization
    \begin{subequations}\label{eq:hybrid:relax1}
\begin{align+}
	\nonumber\max_{\bd c}\,\,& f(\bd c)\\
 \label{subeq:hybrid:relax1:bin}	\mathrm{s.t.}\,\,& c_{i}^2-c_i=0\,\,\,\,\, i=1...MN,\\
 \label{subeq:hybrid:relax1:TotNum}\,& \bd c^{T}\bd c=k_\mt mk_\mt r,\\
     \label{subeq:hybrid:relax1:RxMem}\,\,& \text{card}\left(\bd c^T \bd P_r \right)\leq k_\mt t,\\
    \label{subeq:hybrid:relax1:TxMem}\,\,& \bd c^{T}\bd Q_{r}\bd c=k_\mt m^2 k_\mt r\\
    \label{subeq::hybrid:relax1:addconst2}&\bd c^{T}\bd P_{{\mt t,i}}\bd c\leq k_\mt r\,\,\,\,\, i=1...M,\\
    \label{subeq::hybrid:relax1:addconst1}\,\,&\bd c^{T}\bd P_{{\mt r,i}}\bd c\leq k_\mt m\,\,\,\,\, i=1...N,
    \end{align+}    
    \end{subequations}
where $\bd P_\mt t$ is 
\begin{align}
\bd P_\mt t&=\left[\bd v_{\mt t,1},...,\bd v_{\mt t,M}\right].
\end{align}
The cardinality constraint (\ref{subeq:hybrid:relax1:RxMem}), is a nonconvex constraint. The best relaxation strategy for a cardinality or norm-0 constraint is a norm-1 constraint \cite{boyd2007l1}, which is already met by (\ref{subeq:hybrid:relax1:TotNum}). To tackle this constraint, we employ alternative constraints described in the following theorem.
\begin{theorem}\label{theo:hybrid_alternative_constraints}
Let $\bd c$ be the selection vector associated with the hybrid selection satisfying (\ref{subeq:hybrid:relax1:bin}) to (\ref{subeq:hybrid:relax1:TotNum}). Then, we can replace (\ref{subeq:hybrid:relax1:RxMem}) by the following inequality constraints
\begin{align}
        &\frac{(k_\mt rk_\mt m)^2}{k_\mt t}\leq\bd c^T\bd Q_\mt t \bd c \leq k_\mt r^2k_\mt m,
\end{align}
where $\bd Q_\mt t=\bd P_\mt t\bd P_\mt t^T$.
\end{theorem}
\begin{IEEEproof}
See Appendix \ref{sec:appendix_Proof_theorem4}.
\end{IEEEproof}

We relax (\ref{subeq:hybrid:relax1:RxMem}) through (\ref{subeq::hybrid:relax2:addconst3}), and ultimately reformulate the hybrid selection as follows
    \begin{subequations}\label{eq:hybrid:relax2}
\begin{align+}
	\nonumber\max_{\bd c}\,\,& f(\bd c)\\
 \label{subeq:hybrid:relax2:bin}	\mathrm{s.t.}\,\,& c_{i}^2-c_i=0\,\,\,\,\, i=1...MN,\\
 \label{subeq:hybrid:relax2:TotNum}& \bd c^{T}\bd c=k_\mt mk_\mt r\\
      \label{subeq::hybrid:relax2:addconst3}& \frac{(k_\mt rk_\mt m)^2}{k_\mt t}\leq\bd c^{T}\bd Q_{\mt t}\bd c\leq k_\mt r^2k_\mt m ,\\
 \label{subeq:hybrid:relax2:rec}& \bd c^{T}\bd Q_\mt r\bd c=k_\mt m^2k_\mt r,\\
       \label{subeq::hybrid:relax2:addconst2}&\bd c^{T}\bd P_{{\mt r,i}}\bd c\leq k_\mt m\,\,\,\,\, i=1...N,\\
    \label{subeq::hybrid:relax2:addconst1}&\bd c^{T}\bd P_{{\mt t,i}}\bd c\leq k_\mt r\,\,\,\,\, i=1...M.
    \end{align+}    
    \end{subequations}
In the following theorem, we study the performance of the hybrid selection with respect to the rest of the modes.
    \begin{theorem}\label{theo:all}
Let $f^{\text{fct}}$ be the optimal value of the factored selection (\ref{eq:fct}), $f^{\text{hyb}}$ the MFC selection (\ref{eq:mfc}), and $f^{\text{hyb}}$ the optimal value of the hybrid selection (\ref{eq:hybrid:relax2}). Also, let $k_\mt r^{\mt{fct}}$, $k_\mt t^{\mt{fct}}$ be the number of selected transmitters and receivers in factored selection, $k_\mt m^{\mt{mfc}}$, $k_\mt r^{\mt{mfc}}$ the number of selected matched filters and receivers in MFC selection, and $k_\mt m^{\mt{hyb}}$, $k_\mt r^{\mt{hyb}}$, $k_\mt t^{\mt{hyb}}$ the number of selected matched filters, receivers, and the number of allowed transmitters in hybrid selection. Moreover, suppose that $k$ elements are selected in each selection strategy subject to the following conditions
\begin{align}
A:\;\;&k=k_\mt r^{\mt{fct}}k_\mt t^{\mt{fct}}=k_\mt m^{\mt{mfc}}k_\mt r^{\mt{mfc}}=k_\mt m^{\mt{hyb}}k_\mt r^{\mt{hyb}}\\
B:\;\;&k_\mt r^{\mt{fct}}=k_\mt r^{\mt{mfc}}=k_\mt r^{\mt{hyb}}\\
C:\;\;&k_\mt t^{\mt{fct}}=k_\mt m^{\mt{mfc}}=k_\mt m^{\mt{hyb}}\\
D:\;\;&k_\mt m^{\mt{hyb}}\leq k_\mt t^{\mt{hyb}}
\end{align}
Then,
\begin{equation}
f^{\mt{fct}} \leq f^{\mt{hyb}}\leq f^{\mt{mfc}}
\end{equation}
\end{theorem}
\begin{IEEEproof}
We can rewrite (\ref{eq:fct}) as 
\begin{subequations}
\begin{align+}\label{eq:fct:expanded}
	\max_{\bd c}\,\,& f(\bd c)\\
	 \mathrm{s.t.}\\
	\mu_i&: \;\; c_{i}^2-c_i=0\,\,\,\,\, i=1...MN,\\
		 \nu&:\;\; \bd c^{T}\bd c=k_\mt tk_\mt r\\
	\lambda&:\;\; \bd c^{T}\bd Q_\t{r}\bd c= k_\mt t^2k_\mt r\\
	 \kappa&:\;\; \bd c^{T}\bd Q_\t{t}\bd c=k_\mt tk_\mt r^2\\
	\rho_i&: \bd c^{T}\bd P_{{\mt t,i}}\bd c\leq k_\mt r \,\,\,\,\, i=1...M,\\
		\label{eq:fct_Rx} \eta_i&: \bd c^{T}\bd P_{{\mt r,i}}\bd c\leq k_\mt t \,\,\,\,\, i=1...N,
	\end{align+}
	\end{subequations}
while, we revise (\ref{eq:hybrid}) as:
\begin{subequations}\label{eq:hyb:expanded}
\begin{align+}
	\nonumber\max_{\bd c}\,\,& f(\bd c)\\
	 \mathrm{s.t.}\\
	\mu_i&: \;\; c_{i}^2-c_i=0\,\,\,\,\, i=1...MN,\\
		 \nu&:\;\; \bd c^{T}\bd c=k_\mt tk_\mt r\\
	\label{eq:hyb_PP_T} \lambda&:\;\; \bd c^{T}\bd Q_\t{r}\bd c= k_\mt t^2k_\mt r\\
	\label{eq:hyb_PP_r} \kappa&:\;\; \bd c^{T}\bd Q_\t{t}\bd c= k_\mt tk_\mt r^2\\
		 \sigma&:\;\; \bd c^{T}\bd Q_\t{t}\bd c < k_\mt r^2k_\mt t\\
	\label{eq:hyb_Q_r_nc}	 \tau&:\;\; \bd c^{T}\bd Q_\t{t}\bd c\ge \frac{(k_\mt rk_\mt m)^2}{k_\mt t}\\
	\label{eq:hyb_Tx}\rho_i&: \bd c^{T}\bd P_{{\mt t,i}}\bd c\leq k_\mt r \,\,\,\,\, i=1...M,\\
		\label{eq:hyb_Rx} \eta_i&: \bd c^{T}\bd P_{{\mt r,i}}\bd c\leq k_\mt m \,\,\,\,\, i=1...N.
	\end{align+}
	\end{subequations}
	By computing the Lagrangian dual function of (\ref{eq:fct:expanded}), and (\ref{eq:hyb:expanded}) we get
	\begin{align}
	    g^{\mt{hyb}}-g^{\mt{fct}}= -\frac{1}{4}\bs\omega^T\left(\left(\sigma+\tau\right)\bd Q_\mt t\right)\bs\omega-\sigma k_\mt r^2k_\mt t-\tau \frac{(k_\mt rk_\mt m)^2}{k_\mt t}. 
	\end{align}
Due to the KKT conditions, we note that
\begin{align}
     \sigma \geq 0,\;\; \tau \geq 0.
\end{align}
Therefore,
	\begin{align}
	    g^{\mt{hyb}}-&g^{\mt{fct}}\leq 0,
	\end{align}	
and consequently
	\begin{align}
	    f^{\mt{fct}} &\leq f^{\mt{hyb}}.
	\end{align}
Following the same process for comparing $f^{\mt{hyb}}$ and $f^{\mt{mfc}}$ yields
	\begin{align}
	    f^{\mt{hyb}} &\leq f^{\mt{mfc}}.
	\end{align}
\end{IEEEproof}


\subsection{Performance and Complexity Analysis}
\begin{figure}[!bt]
    \centering
    \includegraphics[width=5.5cm]{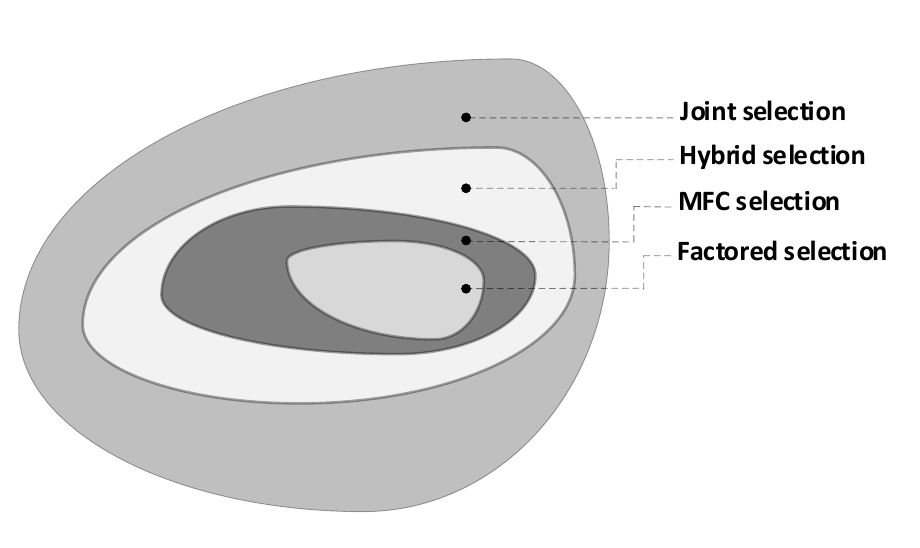}
    \caption{Geometric illustration of feasible regions and relationship of different selection methods.}
    \label{fig:feasible_set}
\end{figure}

By Theorem \ref{theo:mfc_joint} , and Theorem \ref{theo:all}, we conclude the following, in terms of the output $\mt{SINR}$,
\begin{align}
f^{\mt{fct}} \leq f^{\mt{hyb}}\leq f^{\mt{mfc}}\leq f^{\mt{jnt}},
\end{align}
Also, based on this conclusion the feasible regions for different selection modes given by intersection of positive semidefinite cones can be illustrated as in Fig. \ref{fig:feasible_set}.

The computational complexity of the minimum volume ellipsoid problem  when solving via barrier-generated path-following interior-point method with $n$ variables and $m$ constraints can be approximated by $O\left(m^{2.5}(n^2+m)\right)$ (see Theorem 6.5.1 in \cite{Nesterov1994}). We can represent the joint selection with at least $m^\mt{jnt}=2$ constraints. However, the factored and MFC selections need at least $m^\mt{fct}=m^\mt{mfc}=3+M+N$ constraints, and the hybrid selection requires $m^\mt{fct}=m^\mt{mfc}=5+M+N$ constraints. Hence, the computational complexities can be summarized as
\begin{align}
\nonumber O^{\mt{jnt}}\leq O^{\mt{fct}} = O^{\mt{mfc}} \leq O^{\mt{hyb}}.
\end{align}
 \section{Relaxation and Approximation}\label{sec:relaxation}   
In the previous section, we expressed the selection methods as a difference log-determinant maximization with quadratic constraints. 
Although we decoupled the factored problem in subsection \ref{subsec:factored}, and relaxed the norm constraints in  subsection \ref{subsec:hybrid}, the problem still is not tractable due to the nonconvexity of (a) a difference of concave functions, (b) quadratic equalities, e.g. (\ref{eq:fct_bin})- (\ref{eq:fct_fct}), and (c) the quadratic inequality from constructing a nonconvex set, e.g. (\ref{eq:hyb_Q_r_nc}). To handle these nonconvexities, we propose a variation of sequential convex programming (SCP) with an exact penalty approach \cite{Lipp2015}. Let the primal problem be defined as  
\begin{align}
	\nonumber\max_{\bd c}\,\,&f(\bd c)\\
	\nonumber\mathrm{s.t.}\,\,&  c_{i}\in \left\lbrace0,1\right\rbrace\,\,\,\,\, i=1...MN,\\
	\nonumber	\,\,&  g_1(\bd c)\geq 0\\
	\nonumber\,\,&  g_i(\bd c)\leq 0,\;\;\;\;i=1,...,\ell,\\
	\,\,&  e_i(\bd c)= 0,\;\;\;\;i=1,...,\jmath.
\end{align}
Where $f(\bd c)$ is a concave-concave function, $g(\bd c)$, and $e(\bd c)$ represent quadratic functions. By relaxing the binary constraint and using a convex local approximation we specify the problem in the $\ell$-th iteration as
\begin{align}\label{eq:scp}
	\max_{\bd c^{(l)}}\,\,&\hat{f}(\bd c^{(l)})+\psi \sum\limits_{i=1}^{\jmath} e_i(\bd c^{(l)})\\
	\nonumber\mathrm{s.t.}\,\,&  0 \leq c^{(i)}_{i} \leq 1\,\,\,\,\, i=1...MN,\\
	\nonumber	\,\,&  \hat{g}_1(\bd c^{(l)})\geq 0\\
	\,\,&  g_i(\bd c^{(l)})\leq 0,\;\;\;\;i=1,...,\ell.
\end{align}
Where we use a first Taylor expansion to get an affine approximation over the trust region of a box around the current point. Employing this for $f(\bd c)$ we will have 
\begin{align}
        \nonumber\hat{f}(\bd c^{(l)})&=f_1({\bd c^{(l)}})-f_2({\bd c^{(l)}})-\triangledown f_2({\bd c^{(l)}}) (\bd c^{(l)}-{\bd c^{(l-1)}}) \\
       \nonumber &=\mt{log det}\left(\bd A_{\mt{s}}^H \mt{diag}(\bd c^{(l)}) \bd A_{\mt{s}}+\bd B_{\mt{s}}\right)\\
       \nonumber &-\mt{log det}\left(\bd A_{\mt{jc}}^H \mt{diag}(\bd c^{(l-1)}) \bd A_{\mt{jc}}+\bd B_{\mt{jc}}\right)\\
       &-\mt{diag}\left(\bd a_s^H \left(\bd A_{\mt{jc}}^H \mt{diag}\left(\bd c^{(l-1)}\right) \bd A_{\mt{jc}}+\bd B_{\mt{jc}}\right)^{-1} \bd a_s \right).
\end{align}
We also use an affine approximation for $g_1(\bd c)$ to tackle the nonconvex constraint
\begin{align}
  \nonumber&\hat{g}_1(\bd c^{(k)})= g_1(\bd c^{(k)})+\triangledown g_1({\bd c^{(k)}}) (\bd c^{(k)}-{\bd c^{(k-1)}}\\
  \nonumber&= {\bd c^{(k)}}^T \bd W_1 \bd c^{(k)}+r_1+ \left(\left(\bd W_1+\bd W_1^T\right)\bd c^{(k)}\right)\left(\bd c^{(k)}-{\bd c^{(k-1)}}\right)
\end{align}
After solving the relaxed version of the problem by an SCP procedure, we find the suboptimal solution of the original problem by an appropriate randomized rounding strategy. 



Let ($ \bd c^\star$) be the optimal value obtained by relaxation and SCP. Also, assume $\bd z \in \bd {R}^{MN}$ is a Gaussian variable with distribution $\bd z \sim \mathcal{N}(\mu,\Sigma)$. Then, the following maximization
\begin{align}
	\nonumber\max_{\bd c}\,\,&\mathbb{E}\left(f(\bd c)\right)\\
	\nonumber\mathrm{s.t.}\,\,& \mathbb{E}\left( c_{i}\right)\in \left\lbrace0,1\right\rbrace\,\,\,\,\, i=1...MN,\\
	\nonumber	\,\,&  \mathbb{E}\left(g_1(\bd c)\right)\geq 0\\
	\nonumber\,\,& \mathbb{E}\left( g_i(\bd c)\right)\leq 0,\;\;\;\;i=1,...,\ell,\\
	\,\,&  \mathbb{E}\left(h_i(\bd c)\right)= 0,\;\;\;\;i=1,...,\jmath,
\end{align}
is solved by $\bd z$ for $\mu=\bd c^\star$. As for $\Sigma$,  we use $\Sigma=\mt{diag}\left(\left[ \mt{var} (\bd c_i)\right]\right)$ where $\bd c_i$ is a vector comprising the sequence of solutions of (\ref{eq:scp}).
We take a sample from $\bd z$ for a sufficent number of times and and keep the best sample yielding the maximum value.
Nevertheless, this direct sampling does not immediately provide a feasible point regarding the embedded quadratic constraints. Hence, we need to project the direct sampled vector onto the constraints feasible set. To do this we add one more step after sampling, where we find the projected point $\hat{\bd z}$ via the following optimization
\begin{align}\label{eq:optimal_rounding}
  \underset{\hat{\bd z}}{\mathrm{min}}\;\; & \|\hat{\bd z}-\bd z\|\\
  \nonumber\mathrm{s.t.}\,\,&  c_{i}\in \left\lbrace0,1\right\rbrace\,\,\,\,\, i=1...MN,\\
\nonumber		\,\,&  g_1(\bd c)\geq 0\\
\nonumber	\,\,&  g_i(\bd c)\leq 0,\;\;\;\;i=1,...,\ell,\\
	\,\,&  h_i(\bd c)= 0,\;\;\;\;i=1,...,\jmath.
\end{align}
This optimization is another form of binary programming. Heuristically, we approximate this problem in two successive steps. We first find the projected sample by applying a relaxation as follows
\begin{align}\label{eq:optimal_rounding}
  \nonumber\underset{\hat{\bd z}}{\mathrm{min}}\;\; & \|\hat{\bd z}-\bd z\|\\
 \nonumber\mathrm{s.t.}\,\,&  0\leq c_{i}\leq 1\,\,\,\,\, i=1...MN,\\
\nonumber	\,\,&  g_i(\bd c)\leq 0,\;\;\;\;i=1,...,\ell,\\
	\,\,&  h_i(\bd c)\leq 0,\;\;\;\;i=1,...,\jmath.
\end{align}
Then the best feasible projected sample $\hat{\bd z}^*$ is obtained through successive evaluations of the cost function. We then employ a structured rounding strategy to round $\hat{\bd z}^*$ to the nearest binary point, while still meeting the structure of the selection vector for different selection methods. The corresponding structured rounding for each selection method is listed in Algorithm \ref{alg:str_rounding}, while the final algorithm is summarized in Algorithm \ref{alg:Constrained_randomized}.
\begin{algorithm}[t]
\label{alg:str_rounding}
    \SetKwInOut{Input}{Input}
    \SetKwInOut{Output}{Output}
    \Switch{Selection method}{
\Case{Joint selection}{
  sort $\hat{\bd z}^*$ in descending order\\
round the first $k^\mt{jnt}$ elements to one and the remaining elements to zero
}
\Case{Factored selection}{
    reshape $\hat{\bd z}^*$ as a MIMO selection matrix \\
    calculate sum of the rows, and columns. Then, sort it in descending order\\
    round the first $k_\mt r^\mt{fct}$ rows (receivers), and  $k_\mt t^\mt{fct}$ columns (transmitters) to one and the rest to zero
}
\Case{MFC selection}{
    reshape $\hat{\bd z}^*$ as a MIMO selection matrix \\
    calculate sum of the rows, and columns. Then, sort it in descending order \\
    sort rows descending\\
    round the first $k_\mt m^\mt{mfc}$ elements in the first $k_\mt r^\mt{mfc}$ rows (receivers), and  $k_\mt t^\mt{fct}$ columns (transmitters) to one and the rest to zero
}
\Case{Hybrid selection}{
    reshape $\hat{\bd z}^*$ as a MIMO selection matrix \\
    calculate sum of the columns. Then, sort it in descending order \\
    sort rows descending\\
    round the first $k_\mt m^\mt{mfc}$ elements in the first $k_\mt r^\mt{mfc}$ rows (receivers), to one and the rest zero
}
}
\caption{\small Structured binary rounding } \label{alg:MIMO_FIIB}
\end{algorithm}
\begin{algorithm}[!b]
\caption{\small Optimal Randomized Algorithm } \label{alg:Constrained_randomized}
\small Solve SCP relaxed problem (\ref{eq:scp}) to get $\bd c^\star$\\
\small Initialize the best point $\hat{\bd z}^*:=0$ and $f^{*}:=0$\\
\For{each iteration $n$}{
{\small Random sampling $\bd z \sim \mathcal{N}(\mu,\Sigma)$}\label{alg:Constrained_randomized_S4}\\
{\small Solve (\ref{eq:optimal_rounding}) and get $\hat{\bd z}$  }\\
\If{$f(\hat{\bd z}) \geq f^{\text{best}}$}{$ f^{\text{best}}=f(\hat{\bd z})$ and $\hat{\bd z}^*=\hat{\bd z}$ }
}
\small Employ Algorithm \ref{alg:str_rounding} to obtain $\bd z$
\end{algorithm}

\section{Simulation}\label{sec:sim}
\begin{figure}[!t]
\centering
\subfigure[]{
\includegraphics[width=2.5in]{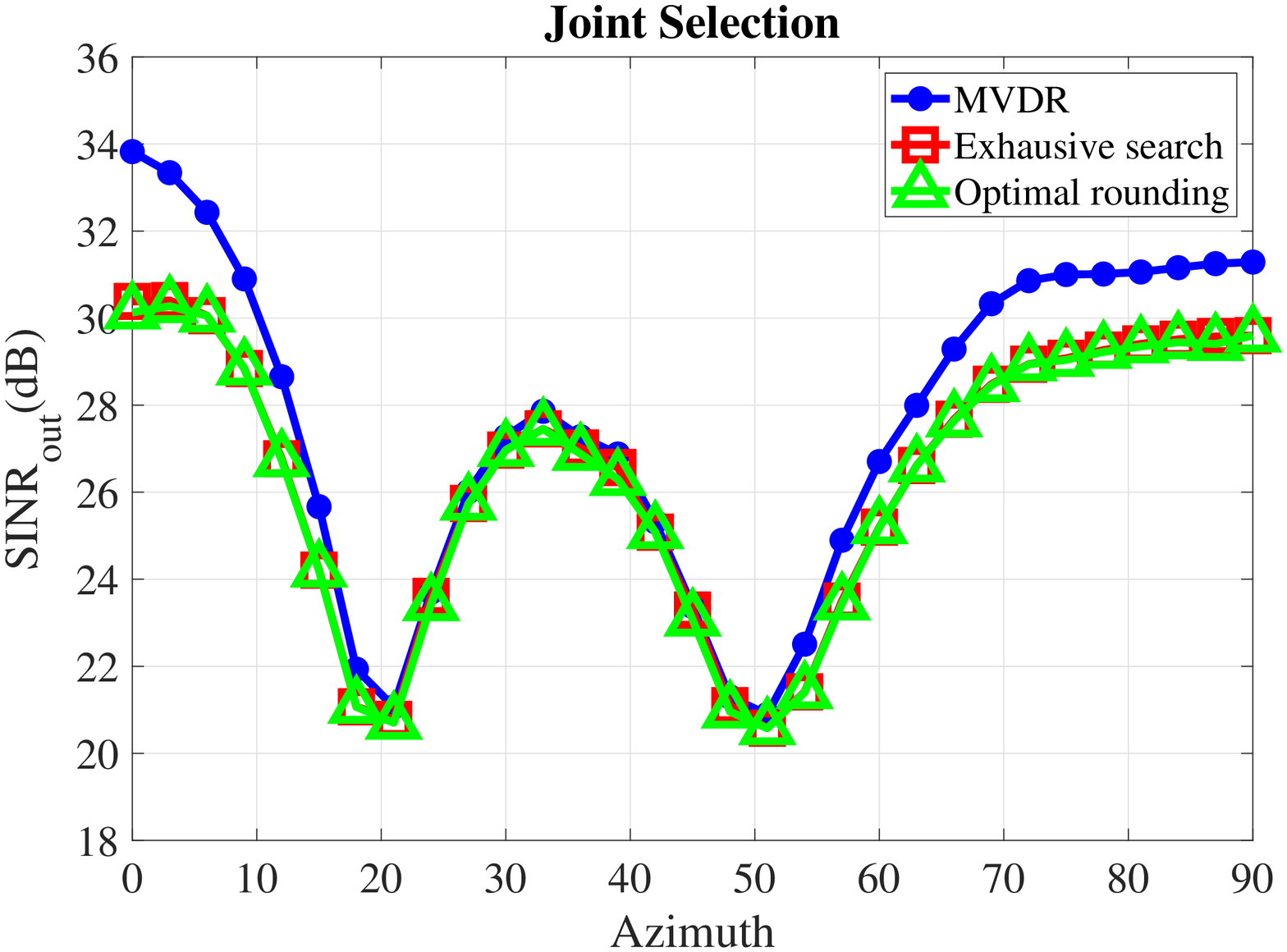}
\label{fig:Joint_bound}
}
\subfigure[]{
\includegraphics[width=2.5in]{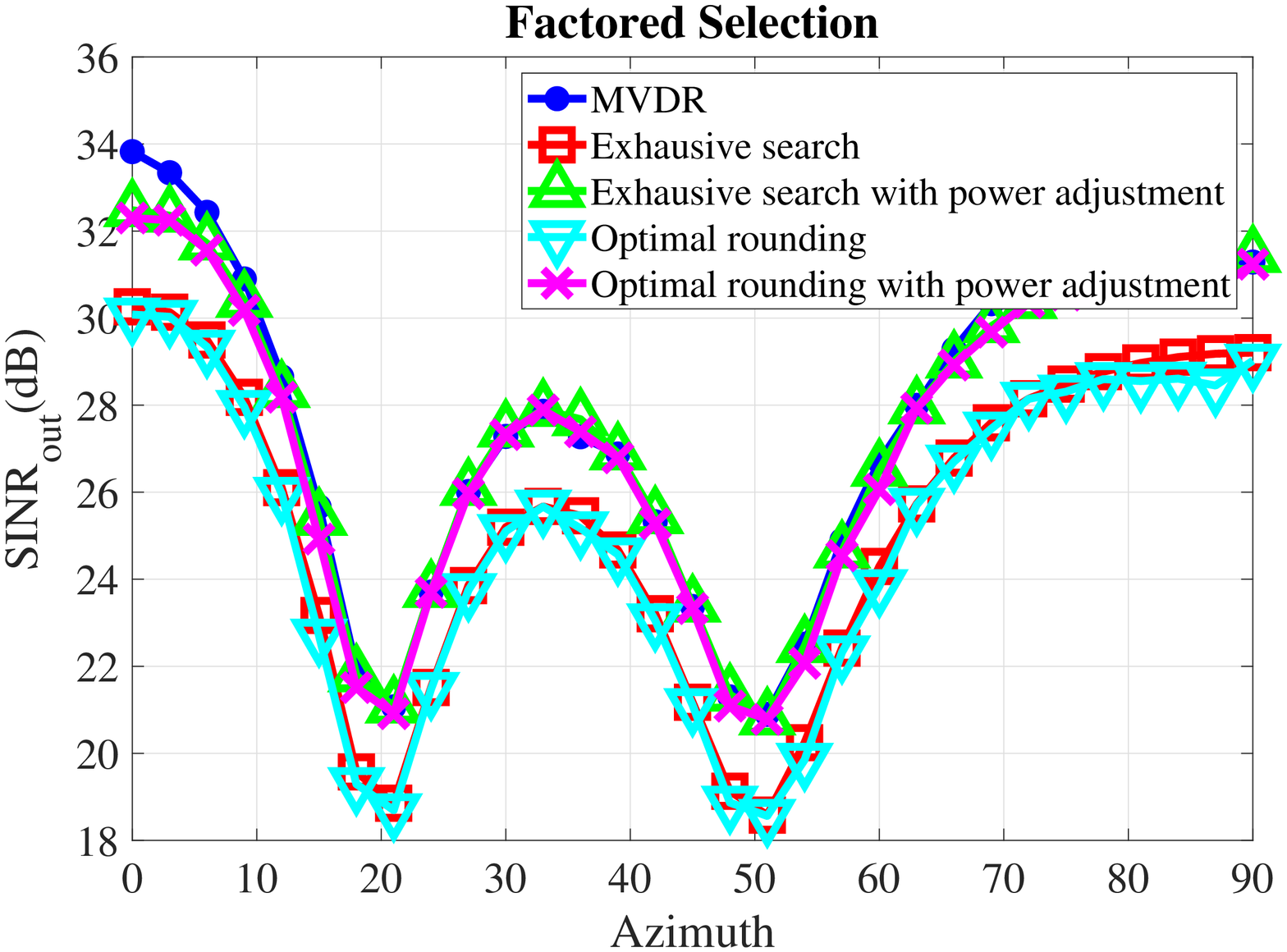}
\label{fig:fct_bound}
}
\subfigure[]{
\includegraphics[width=2.5in]{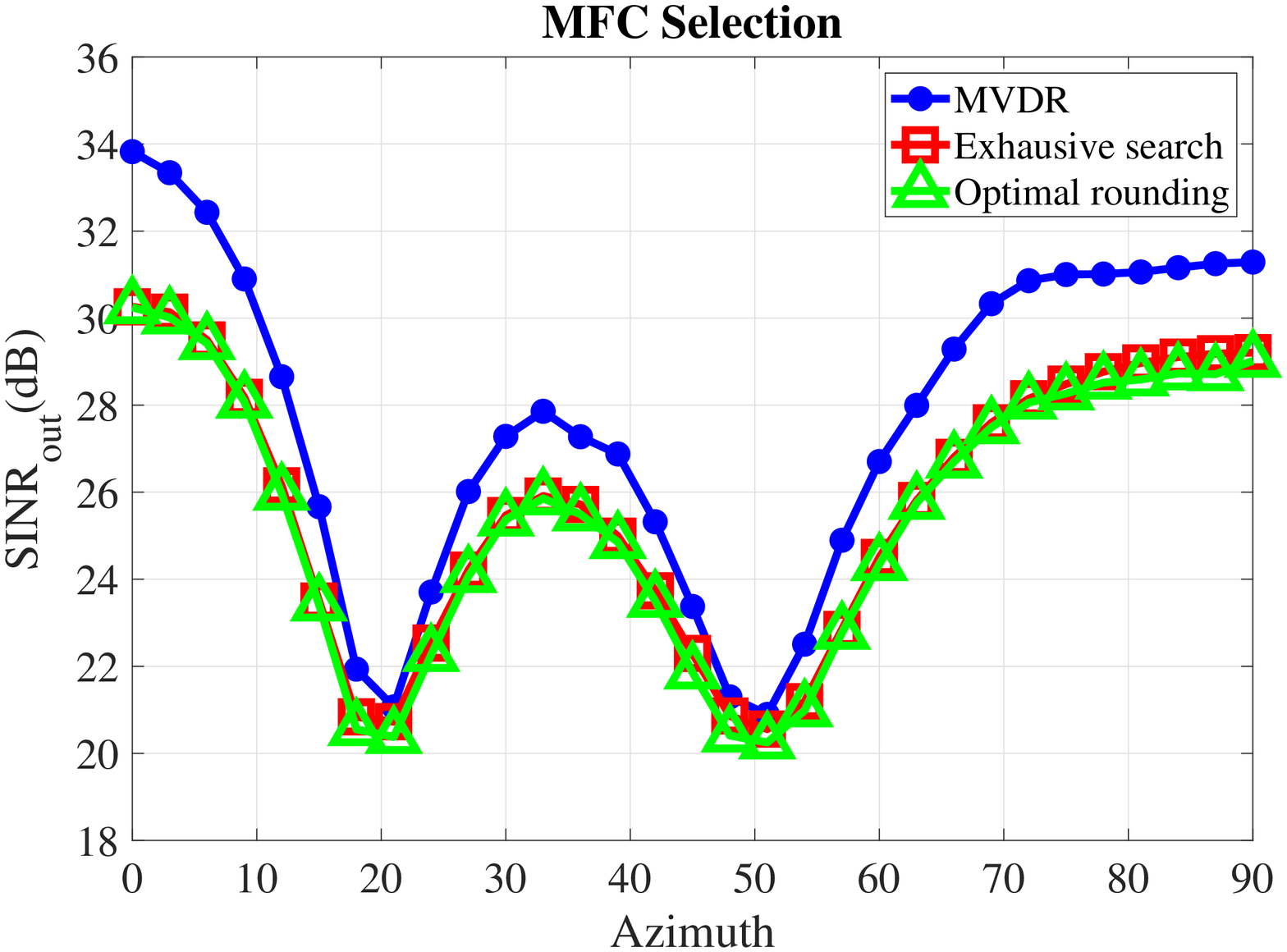}
\label{fig:mfc_bound}
}
\subfigure[]{
\includegraphics[width=2.5in]{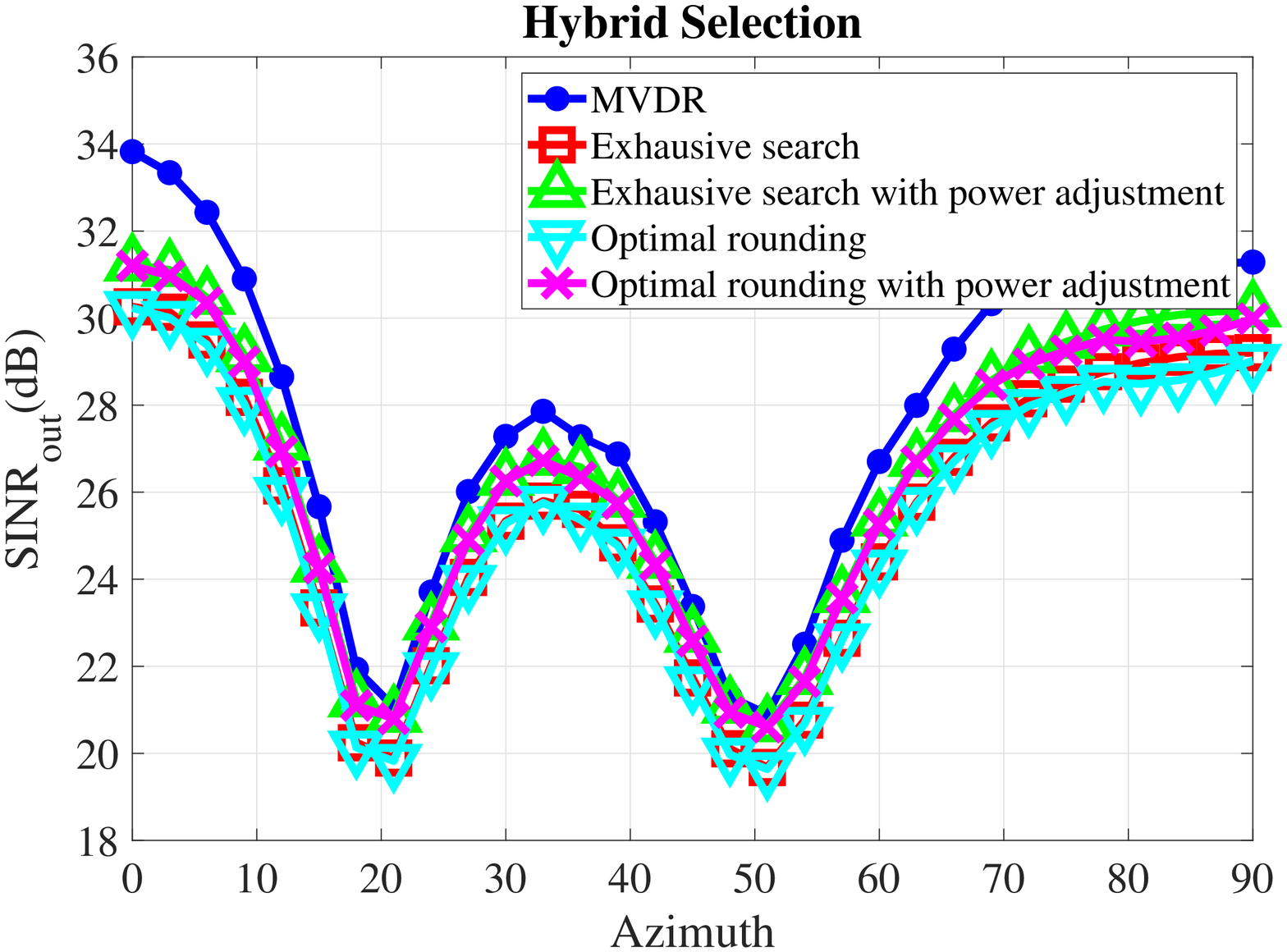}
\label{fig:hyb_bound}
}
\caption{ \label{CooMod} $\mt{SINR}_\mt{out}$ value estimated by MVDR beamformer in the full array vs. the optimum value achieved by exhaustive search, and the proposed optimization for \subref{fig:Joint_bound} Joint selection, \subref{fig:fct_bound} Factored selection, \subref{fig:mfc_bound} MFC selection, and \subref{fig:hyb_bound} Hybrid selection.  }
\end{figure}

To evaluate the performance of the proposed selection methods,  we use a uniform linear collocated MIMO phased array comprising 5 transmitters ($M=5$), and 5 receivers ($N=5$). We use the inter-element spacing of $d_\mt r=\frac{\lambda}{2}$ for the elements in the receive array. To maintain a non-overlapping virtual array and maximum aperture we place the transmit antennas $d_\mt t=Nd_\mt r$ apart. We include two jamming signals with azimuth angles of $\theta_j = [20,50]$ degrees and powers of 13dBW. We vary the azimuth angle of the received signal, $\theta_s$ from $0^\circ$ to $90^\circ$, while the power is fixed at 20 dBW. Also, we assume that clutter is contributed from angles between $0^\circ$  and $90^\circ$, and we use a low rank model with rank 5 and a total clutter-to-noise ratio of 13 dB. We define the problem such that 12 elements are selected in total according to different selection methods, such that for joint selection $k^\mt{jnt}=12$, and for the factored selection $k_\mt t^\mt{fct}=3, k_\mt r^\mt{fct}=4$. The selection criteria for MFC and hybrid selections are $k_\mt m^\mt{mfc}=3, k_\mt r^\mt{mfc}=4$ and  $k_\mt t^\mt{hyb}=4, k_\mt m^\mt{hyb}=3, k_\mt r^\mt{hyb}=4$.
The maximum $\mt{SINR}_\mt{out}$ is first calculated by an exhaustive search for each of the strategies and the suboptimal solution is then obtained from Algorithm \ref{alg:Constrained_randomized}. For each direction, we used the SDPT3 solver embedded in CVX \cite{grant2008cvx} to solve (\ref{eq:scp}) for 10 iterations, and then proceeded to take 1000 samples from a random distribution as described in Algorithm \ref{alg:Constrained_randomized}. 
The achieved values for $\mt{SINR}_\mt{out}$ resulting from the exhaustive search and proposed method are depicted in Figs. \ref{fig:Joint_bound}-\ref{fig:hyb_bound} along with the corresponding  $\mt{SINR}_\mt{out}$ for the full array achieved by an MVDR beamformer. The approximated value is close to the exhaustive search, confirming that a good approximation ratio is achieved by the proposed relaxation strategy.
For the factored and hybrid selection two sets of curves are shown in Figs. \ref{fig:fct_bound}, \ref{fig:hyb_bound} to demonstrate the effect of the power adjustment. The additional transmit power afforded by allocating the total available power to fewer transmit waveforms improves the $\mt{SINR}_{\mt{out}}$ significantly in these modes of selection.  





\begin{figure*}[!bt]
	\centering
		\includegraphics[width=4in]{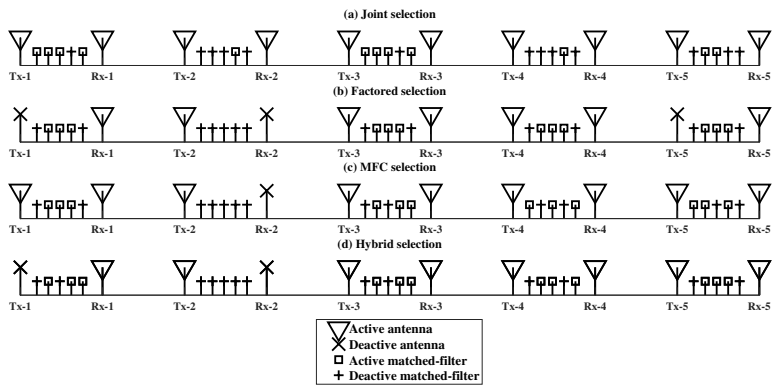}
		\caption{Illustration of the output selection in different selection modes.}
	\label{fig:antennas}
\end{figure*}

Next, we show in Fig.~\ref{fig:antennas} the selected elements yielded by each of the selection strategies. In this example, we fix the impinging signal direction at $\theta_s=18^\circ$. We can see from Fig.~\ref{fig:antennas}(a) that all of the transmitters and receivers must be operational in the joint selection in order to be able to select from the corresponding matched-filters. On the other hand, exactly two transmitters and two receivers were deactivated in factored selection as shown in Fig.~\ref{fig:antennas}(b). For MFC selection three matched filters are selected precisely at four receivers. While in hybrid selection the three matched filters are selected only from 4 possible options as Tx-1 is deactivated.

\begin{figure}[!tb]
	\centering
		\includegraphics[width=2.5in]{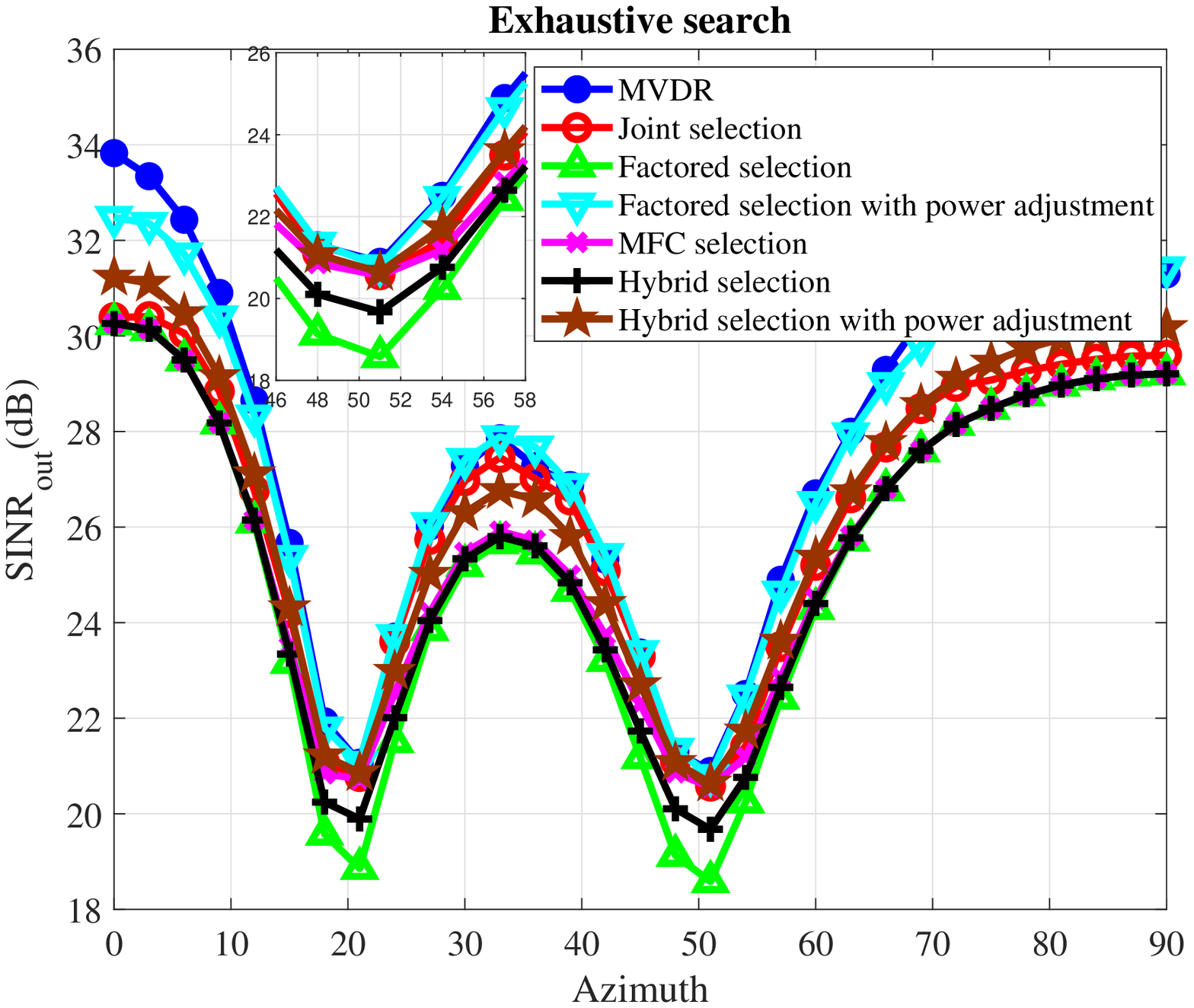}
		\caption{$\mt{SINR}_\mt{out}$ value estimated by MVDR beamformer in the full array vs. the optimum value achieved by exhaustive search in  different selection modes.}
	\label{fig:exh_bound}
	\centering
		\includegraphics[width=2.5in]{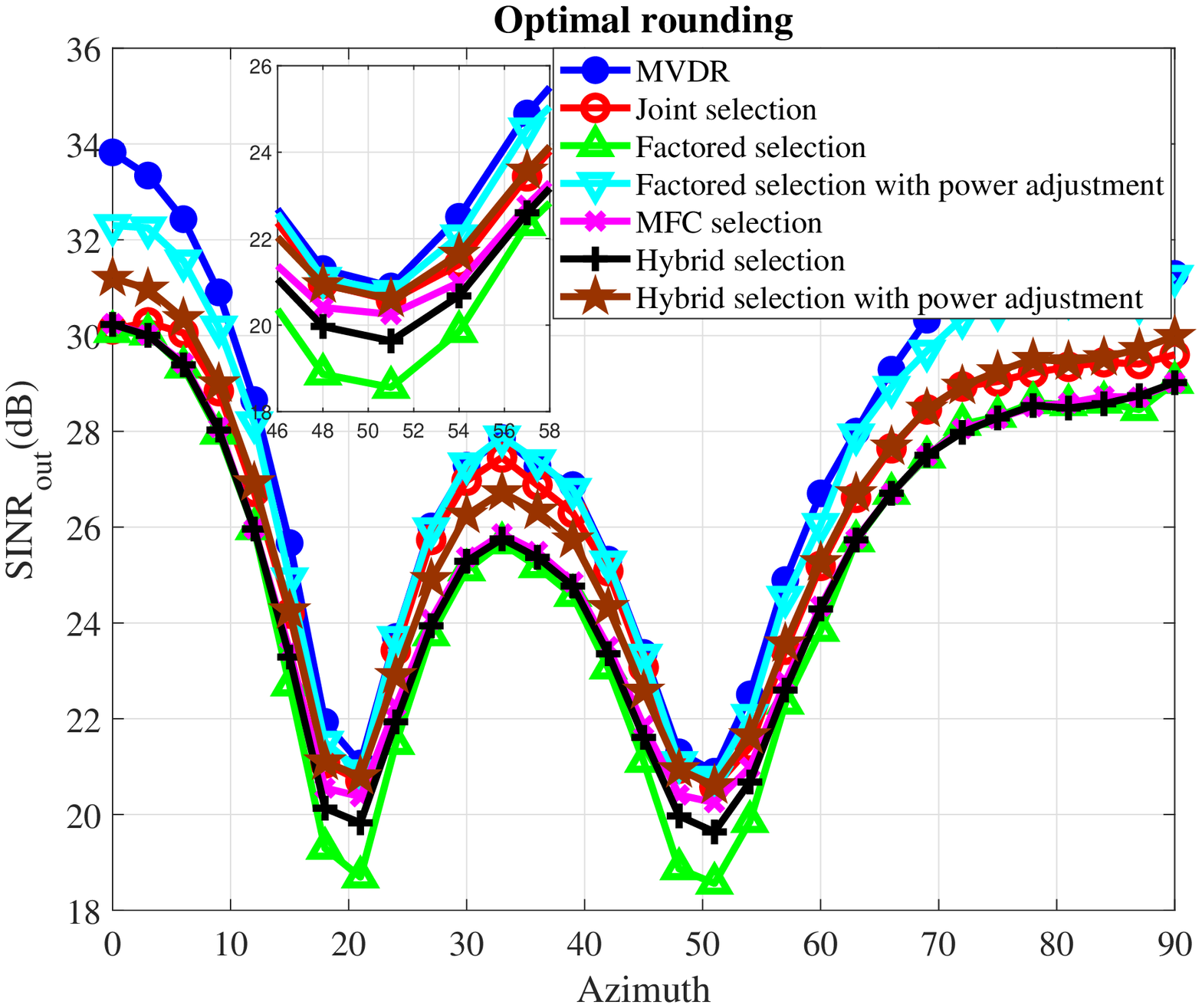}
		\caption{$\mt{SINR}_\mt{out}$ value estimated by MVDR beamformer in the full array vs. the optimum value achieved by proposed optimization in  different selection modes.}
	\label{fig:det_bound}
\end{figure}

We now compare the performances of the selection methods. In  Fig.~\ref{fig:exh_bound} we plot the $\mt{SINR}_\mt{out}$ achieved by the different selection methods via exhaustive search, While in Fig.~\ref{fig:det_bound} the optimal values are shown for the proposed optimization. Ignoring the power adjustment option, we see in both figures that joint selection achieves the highest $\mt{SINR}_\mt{out}$ followed by MFC, hybrid, and then factored selection, which verifies Theorems (\ref{theo:fct}) and (\ref{theo:all}). However, when the power adjustment for factored and hybrid selections is included, the increased transmit signal power means that the factored and hybrid selections surpass the joint mode.

\begin{figure}[!h]
\centering
		\includegraphics[width=2.5in]{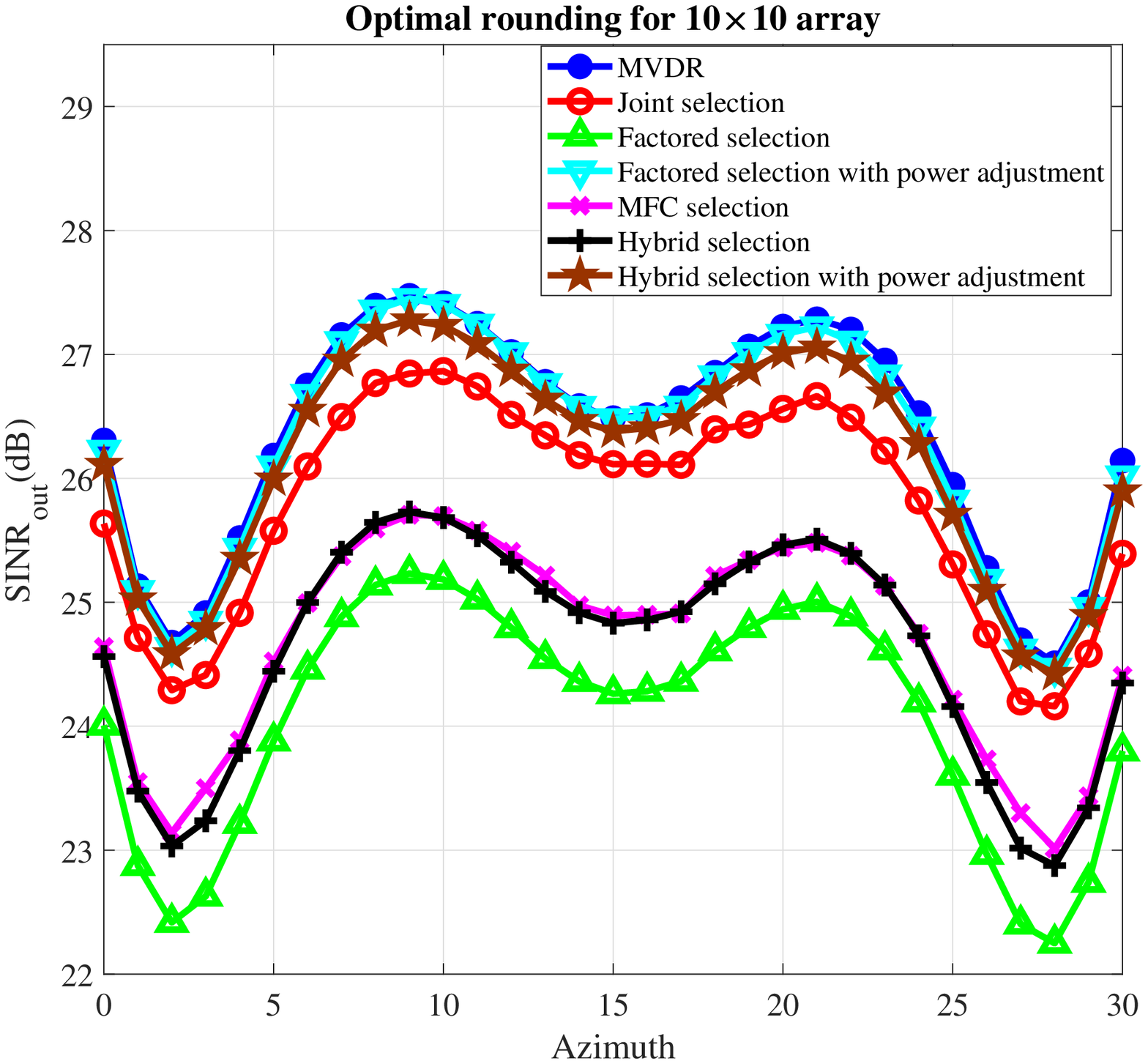}
		\caption{The optimum $\mt{SINR}_\mt{out}$ value achieved by the proposed optimization with varying number of selected elements in different selection modes for a 10 $\times$ 10 array}
	\label{fig:optimal_rounding_bound_10by10}
	\centering
		\includegraphics[width=2.5in]{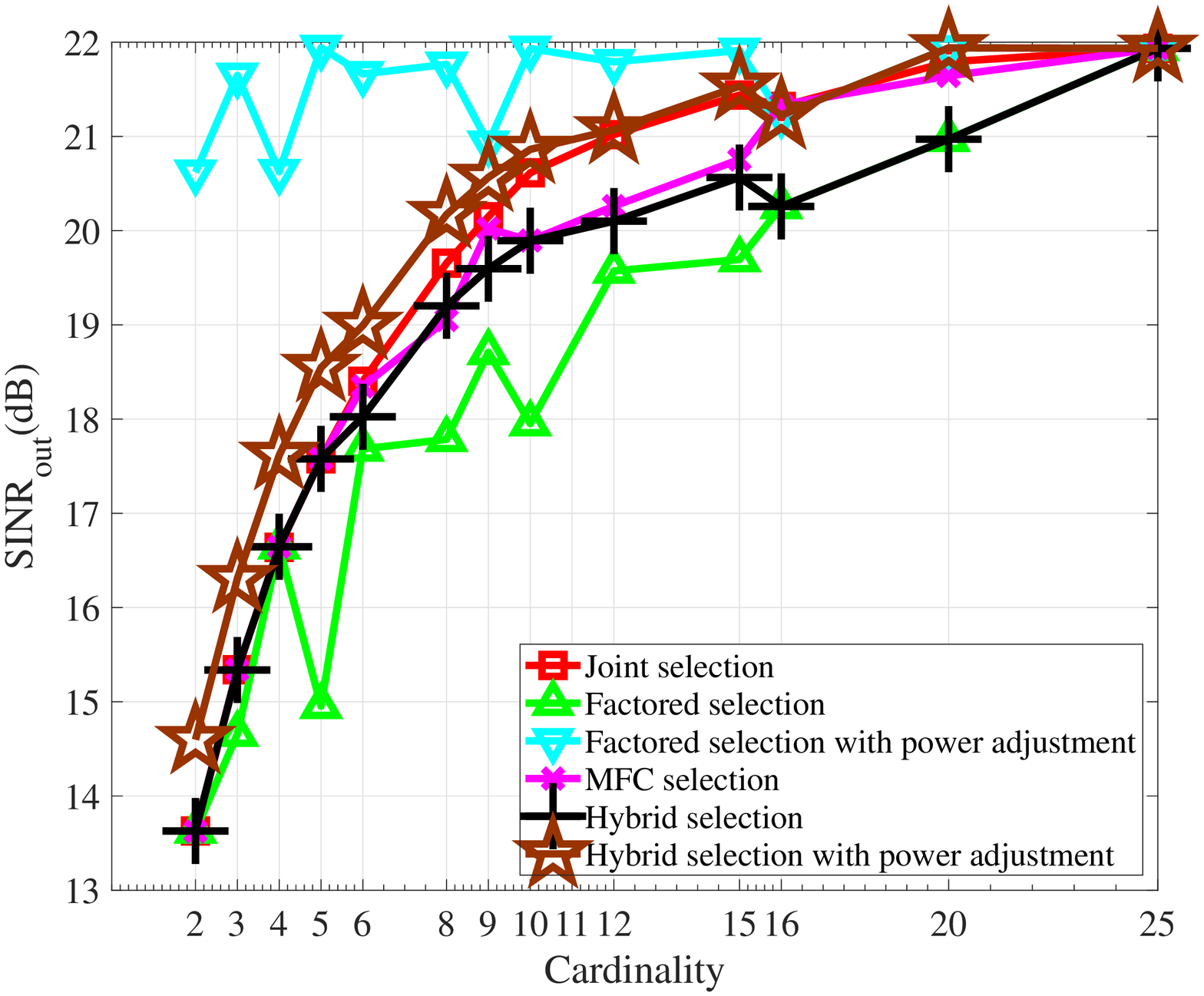}
		\caption{The optimum $\mt{SINR}_\mt{out}$ value achieved by the proposed optimization with varying number of selected elements in different selection modes.}
	\label{fig:bounds_card}
\end{figure}

The above example employed a $5\times 5$ array to permit us to compare the performance to the exhaustive search. Now we demonstrate the performance with a larger array such that the exhaustive search is not possible. To this end, we employ a $10\times 10$ array. Additionally, we increase the number of jammers to 5 and the clutter rank to 10. The five jamming signals have azimuth angles $\theta_j = [0,5,15,25,30]$ degrees and powers of 13dBW, and the total clutter-to-noise ratio is set to 13 dB. The SOI is assumed to have a power of 20dBW at the reciever and its azimuth is varied between $0^\circ$ to $30^\circ$. We select 54 elements in total such that for joint selection $k^\mt{jnt}=54$, and for the factored selection $k_\mt t^\mt{fct}=6, k_\mt r^\mt{fct}=9$. The selection criteria for MFC are $k_\mt m^\mt{mfc}=6, k_\mt r^\mt{mfc}=9$ and for the hybrid strategy $k_\mt t^\mt{hyb}=7,\ k_\mt m^\mt{hyb}=6,\ k_\mt r^\mt{hyb}=9$. The optimal values achieved by the proposed optimization are depicted in Fig. \ref{fig:optimal_rounding_bound_10by10}. Firstly, observe the excellent performance achieved for a larger array with a denser interference environment. The proposed selection strategies exhibit similar trends to the previous, smaller example, with the joint as well as factored and hybrid strategies (with power adjustment) being comparable to the MVDR with almost half the number of elements. 

In the following example, we again use the $5\times 5$ array of the first example. We fix the signal direction $\theta_s= 18^\circ$ and solve the selection problem for an increasing subset of antennas ranging from 2 to 25. The preset values for all the selection methods are listed in Table \ref{table:sim2}. 
As is revealed in Fig.~\ref{fig:bounds_card}, the joint selection outperforms the rest of selection methods when power adjustment is not employed, followed by MFC, hybrid and factored approaches. When the transmit power is adjusted, we see that the factored and hybrid approaches are able to achieve a higher SINR$_{out}$. Also, notice that the output SINR given by the selection remains comparable with that of the full array even when a significantly smaller number of pairs are used. For instance selecting 15 out of 25 elements would substantially reduce the dimensionality and hardware cost but would result in a $\mt{SINR}_\mt{out}$ loss with respect to the full array of less than 0.5 dB for joint selection and 2.24 dB for factored selection (when the power is not adjusted). Adjusting the transmit power, however, can have a significant effect giving, even showing an improvement over the full array.

\begin{table}[!h]
      \caption{The number of selected elements in each selection method for the results shown in Fig. \ref{fig:bounds_card}}
      \label{table:sim2}
      \centering
\begin{tabular}[!t] {|c|c|c|c|c|c|c|c| }
\hline
 $k_\mt{jnt}$ &  $k_\mt t^{\mt{fct}}$  &  $k_\mt r^\mt{fct}$ &  $k_\mt m^\mt{mfc}$ & $k_\mt r^\mt{mfc}$ & $k_\mt t^\mt{hyb}$ & $k_\mt r^\mt{hyb}$&$ k_\mt m^\mt{hyb}$ \\  \hline
2    &     1    &      2    &      1    &    2      &   4      &    2      &    1  \\
	   \hline                                                           
      3    &     1    &      3    &      1    &    3      &   4      &    3      &    1  \\
	  \hline                                                            
      4    &     2    &      2    &      2    &    2      &   4      &    2      &    2  \\
	  \hline                                                            
      5    &     1    &      5    &      1    &    5      &   4      &    5      &    1  \\
	  \hline                                                            
      6    &     2    &      3    &      2    &    3      &   4      &    3      &    2  \\
	  \hline                                                            
      8    &     2    &      4    &      2    &    4      &   4      &    4      &    2  \\
	  \hline                                                            
      9    &     3    &      3    &      3    &    3      &   4      &    3      &    3  \\
	  \hline                                                            
     10    &     2    &      5    &      2    &    5      &   4      &    5      &    2  \\
	 \hline                                                             
     12    &     3    &      4    &      3    &    4      &   4      &    4      &    3  \\
	 \hline                                                             
     15    &     3    &      5    &      3    &    5      &   4      &    5      &    3  \\
	 \hline                                                             
     16    &     4    &      4    &      4    &    4      &   4      &    4      &    4  \\
	 \hline                                                             
     20    &     4    &      5    &      4    &    5      &   4      &    5      &    4  \\
	 \hline                                                             
     25    &     5    &      5    &      5    &    5      &   5      &    5      &    5  \\
	 \hline
\end{tabular}
\end{table}

\section{Conclusion}
\label{sec:conclusion}

In this paper, we formulated the antenna selection in MIMO arrays with beamforming to mitigate interference signals in the form of multiple jamming signals and clutter. We devised four different selection approaches to control different aspects of a MIMO system. We cast the problem as a determinant maximization problem with quadratic constraints to decouple the problem and optimize three joint vectors separately in a unified problem. Since the selection strategies return the optimal subarrays for the given scenario compared to any arbitrary subarray, the maximum output SINR is always achieved. We then presented a theoretical study demonstrating that the joint selection gives the optimum solution followed by the suboptimal solutions offered by MFC, hybrid and factored modes respectively. Moreover, we proposed an appropriate relaxation and approximation method to tackle the nonconvexity of the primal problem. Finally, we presented extensive simulations that verified the theoretical findings and confirmed the effectiveness of the proposed techniques in reducing the problem dimensionality while maintaining a performance that is comparable to the full array. 

\appendices 
\section{Proof of Theorem \ref{theo:added_constraints}}
\label{sec:appendix_Proof_theorem1}

\begin{IEEEproof}
For an arbitrary selection vector we can extend $\mathcal S_1$ under $\mathcal S_2$ and $\mathcal S_3$ as 
\begin{align}\label{eq:addtional}
\bd c^T \bd Q \bd c=  \sum_{i=1}^{M+N} \left(\bd c^T\; [\bd P]_i\right)^2
=\sum\limits_{i=0}^{k_\mt t-1}\epsilon_i(k_\mt t-i)^2+\sum\limits_{i=0}^{k_\mt r-1}\zeta_i(k_\mt r-i)^2,
\end{align}
where $[\bd P]_i$ denotes the $i$-th column of $\bd P$, and $\epsilon_i,\; \zeta_i \in \mathbb{Z}^{*}$, are appropriate integers equal to the number of times the corresponding squared value appears. Here $\mathbb{Z}^{*}$ denotes the set of non-negative integers. 

Now, we can  establish a set of equations based on $\mathcal S_1$ to $\mathcal S_4$ and apply (\ref{eq:addtional}) as follows
%
\begin{align+}
&\label{cond1}\mathcal S_1:\;\;\epsilon_0k_\mt t^2+\sum\limits_{i=1}^{k_\mt t-1}\epsilon_i(k_\mt t-i)^2+\zeta_0k_\mt r^2\\
&+\sum\limits_{i=1}^{k_\mt r-1}\zeta_i(k_\mt r-i)^2=k_\mt rk_\mt t^2+k_\mt tk_\mt r^2\\
&\mathcal S_2:\;\;\epsilon_i \leq k_\mt r\;\;\; i=0,...,k_\mt t-1\\
&\mathcal S_3:\;\;\zeta_i \leq k_\mt t\;\;\; i=0,...,k_\mt r-1\\
\label{cond7}&\mathcal S_4:\;\;\sum\limits_{i=0}^{k_\mt t-1}\epsilon_i(k_\mt t-i)=k_\mt tk_\mt r\\
\label{cond8}&\mathcal S_4:\;\;\sum\limits_{i=0}^{k_\mt r-1}\zeta_i(k_\mt r-i)=k_\mt tk_\mt r.
\end{align+}
Given the above set of equations, we need to show that the only non-trivial solution is 
\begin{align}
\epsilon_0 =k_\mt r \;\;\mt{and}\;\; \epsilon_i=0 &\;\;\;\mt{for}\;\;\; i=1,...,k_\mt t-1\\
\zeta_0=k_\mt t \;\;\mt{and}\;\; \zeta_i=0 &\;\;\;\mt{for}\;\;\; i=1,...,k_\mt r-1.
\end{align}
Based on (\ref{cond7}) and (\ref{cond8}) we can write
\begin{align}
\epsilon_0=k_\mt r-\sum\limits_{i=1}^{k_\mt t-1}\epsilon_i\frac{(k_\mt t-i)}{k_\mt t},\;\;\zeta_0=k_\mt t-\sum\limits_{i=1}^{k_\mt r-1}\zeta_i\frac{(k_\mt r-i)}{k_\mt r}.
\end{align}
Then, we extend (\ref{cond1}) as
\begin{align}
\nonumber&(k_\mt r-\sum\limits_{i=1}^{k_\mt t-1}\epsilon_i\frac{(k_\mt t-i)}{k_\mt t})k_\mt t^2+\sum\limits_{i=1}^{k_\mt t-1}\epsilon_i(k_\mt t-i)^2\\
\nonumber&+(k_\mt t-\sum\limits_{i=1}^{k_\mt r-1}\zeta_i\frac{(k_\mt r-i)}{k_\mt r})k_\mt r^2+\sum\limits_{i=1}^{k_\mt r-1}\zeta_i(k_\mt r-i)^2=k_\mt rk_\mt t^2+k_\mt tk_\mt r^2.
\end{align}
Hence, we have
\begin{align}
\label{eq:hold}\sum\limits_{i=1}^{k_\mt t-1}\epsilon_i i \left(i-k_\mt t\right)+\sum\limits_{i=1}^{k_\mt r-1}\zeta_i i\left(i-k_\mt r\right)=0.
\end{align}
Noting that
\begin{align}
\epsilon_i \geq 0,\;\;\zeta_i \geq 0,\;\;i\geq 1,\;\;(i-k_\mt t) < 0,\;\;(i-k_\mt r) < 0,
\end{align} 
It follows easily that (\ref{eq:hold}) holds if and only if
\begin{align}
\epsilon_i=0 \;\;\text{for}\; i=1,...,k_\mt t-1,\;\;\zeta_i=0 \;\;\text{for}\; i=1,...,k_\mt r-1,
\end{align}
and accordingly
\begin{align}
\epsilon_0=k_\mt r \;\;\;,\;\;\;\zeta_0=k_\mt t.
\end{align}
\end{IEEEproof}
\section{Proof of Theorem \ref{theo:hybrid_alternative_constraints}}
\label{sec:appendix_Proof_theorem4}
\begin{IEEEproof}
We introduce the quadratic function  $e(\bd y), \bd y \in R^{k_\mt t}$ as
\begin{align}
    e(\bd y)=\bd c^T\bd Q_\mt t \bd c=\bd c^T \bd P_\mt t \bd P_\mt t^T \bd c= 
    \sum_{i=1}^{M} \left(\bd c^T\; [\bd P_\mt t]_i\right)^2=
    \sum\limits_{i=1}^{k_\mt t} y_i^2.
    \end{align}
To find the lower-bound, we solve the following minimization
\begin{subequations}
\begin{align}
    \min_{\bd y}\;\; &e(\bd y)\\
    \text{s.t.}\;\; &\bd 1_{k_\mt t}^T \bd y=k_\mt rk_\mt m.
\end{align}
\end{subequations}
Solving this problem using the Lagrangian method yields 
\begin{align}
    y_i=\frac{k_\mt rk_\mt m}{k_\mt t},\;\;\;\; i=1,...,k_\mt t
\end{align}
and subsequently the lower-bound is given by
\begin{align}
e(\bd y)=\frac{(k_\mt rk_\mt m)^2}{k_\mt t}.
\end{align}
The upper-bound is achieved by letting $k_\mt m=k_\mt t$ yielding
\begin{align}
e(\bd y)=k_\mt r^2k_\mt m.
\end{align}
\end{IEEEproof}
\bibliographystyle{IEEEtran}
\balance
\bibliography{IEEEabrv,ref.bib}

\end{document}